\documentclass[aps,prb,showpacs,floatfix,superscriptaddress,twocolumn]{revtex4}

\usepackage{amsmath,amssymb,graphicx}
\usepackage{xcolor}
\usepackage[colorlinks=true,urlcolor=blue,linkcolor=blue,citecolor=blue,bookmarks=false]{hyperref}

\begin{document}
	
\title{Shaking and pushing skyrmions: Formation of a non-equilibrium phase\\
with zero critical current}

\author{Felix Rucker}
 \affiliation{School of Natural Sciences, Technical University of Munich, D-85748 Garching, Germany}

\author{Alla Bezvershenko}
 \affiliation{Institute for Theoretical Physics, Universit\"at zu K\"oln, D-50937 K\"oln, Germany}

\author{Denis Mettus}
 \affiliation{School of Natural Sciences, Technical University of Munich, D-85748 Garching, Germany}
 
\author{Andreas Bauer}
 \affiliation{School of Natural Sciences, Technical University of Munich, D-85748 Garching, Germany}
 \affiliation{Zentrum f\"ur QuantumEngineering (ZQE), Technical University of Munich, D-85748 Garching, Germany}
 
\author{Markus Garst}
\affiliation{Institut f\"ur Theoretische Festk\"orperphysik, Karlsruhe Institute of Technology, D-76131 Karlsruhe, Germany}
\affiliation{Institute for Quantum Materials and Technology, Karlsruhe Institute of Technology, D-76131 Karlsruhe, Germany}
 
\author{Achim Rosch}
 \affiliation{Institute for Theoretical Physics, Universit\"at zu K\"oln, D-50937 K\"oln, Germany}

\author{Christian Pfleiderer}
\affiliation{School of Natural Sciences, Technical University of Munich, D-85748 Garching, Germany}
\affiliation{Zentrum f\"ur QuantumEngineering (ZQE), Technical University of Munich, D-85748 Garching, Germany}
\affiliation{Munich Center for Quantum Science and Technology (MCQST), Technical University of Munich, D-85748 Garching, Germany} 
\affiliation{Munich Center for Quantum Science and Technology (MCQST), Technical University of Munich, D-85748 Garching, Germany} 
\affiliation{Heinz Maier-Leibnitz Zentrum (MLZ), Technical University of Munich, D-85748 Garching, Germany}

\date{\today}

\begin{abstract}	
In three-dimensional chiral magnets, skyrmions are line-like objects oriented parallel to the applied magnetic field. The efficient coupling of magnetic skyrmion lattices to spin currents and magnetic fields permits their dynamical manipulation. Here, we explore the dynamics of skyrmion lattices when slowly oscillating the field direction by up to a few degrees on millisecond timescales while simultaneously pushing the skyrmion lattice by electric currents. The field oscillations induce a shaking of the orientation of the skyrmion lines, leading to a phase where the critical depinning current for translational motion vanishes. We measure the transverse susceptibility of MnSi to track various depinning phase transitions induced by currents, oscillating fields, or combinations thereof. An effective slip--stick model for the bending and motion of the skyrmion lines in the presence of disorder explains main features of the experiment and predicts  the existence of several dynamical skyrmion lattice phases under shaking and pushing representing new phases of matter far from thermal equilibrium.
\end{abstract}

\maketitle

\section{Introduction}

Collective dynamical patterns are ubiquitous in nature, appearing in electronically ordered phases, chemical systems, biological cells, neural networks, and even social contexts. In magnetic materials a particularly rich variety of static patterns has been reported such as conventional magnetic domains~\cite{1998_Hubert_Book}, quantum order~\cite{2017_Keimer_NatPhys, 2021_Giustino_JPhysMater, 2021_Cava_ChemRev}, and topological textures such as skyrmions, merons, and hopfions~\cite{2013_Nagaosa_NatNanotechnol, 2021_Gobel_PhysRep}. This raises the question for the existence and potential properties of collective dynamical patterns  amongst the vast range of magnetic systems which are amenable for theoretical and experimental exploration.

The criteria to be met by a magnetic model system of collective dynamical patterns comprise well-understood forces underlying their formation, and a well-developed theoretical framework providing meaningful insights on the dynamical response. Topological spin textures meet these requirements in an ideal manner. Along with the initial discovery of skyrmion lattices in cubic chiral magnets~\cite{2009_Muhlbauer_Science, 2010_Yu_Nature}, several key characteristics were reported entailing a search for dynamical patterns. In these materials skyrmions, (i), stabilize parallel to an applied field except for tiny corrections by magnetic anisotropies~\cite{2017_Bannenberg_PhysRevB, 2018_Adams_PhysRevLett}, (ii) feature an exceptionally strong coupling to spin currents as inferred initially from ultra-small critical current densities, $j_c\approx 10^6\,\mathrm{Am}^{-2}$~\cite{2010_Jonietz_Science, 2012_Schulz_NatPhys, 2012_Yu_NatCommun}, (iii), display a tiny coupling to disorder and defects as reflected also in the low values of $j_c$~\cite{2011_Zang_PhysRevLett, 2011_Everschor_PhysRevB, 2013_Iwasaki_NatCommun}, and, (iv), are remarkably well-described using phenomenological modeling~\cite{2016_Seidel_Book, 2016_Seki_Book}.

It was, hence, long known that skyrmion lattices in the presence of a current feature two qualitatively different states. First, a static phase where skyrmions do not move, up to corrections from thermal creep~\cite{2013_Lin_PhysRevB, 2020_Luo_CommunMater}, and, second, a dynamic phase in which skyrmions are unpinned and flow with a finite average velocity. However, considering that skyrmions in cubic chiral magnets were found to align parallel to the applied magnetic field to minimize their free energy, it was appreciated that changes of the direction of the magnetic field offer an alternative way to manipulate skyrmion strings. Indeed, so-called time-involved small-angle neutron scattering (TISANE) demonstrated the unpinning of the skyrmion lattice under slow, periodic changes of field orientation above a tiny critical angle of ${\sim}0.4^{\circ}$~\cite{2016_Muhlbauer_NewJPhys, 2022_Mettus_JApplCrystallogr}. This raises the question addressed in our study, if dynamical skyrmion lattice phases can be accessed and created when combining the effects of pushing by spin currents with shaking the orientation of the skyrmion lattices dynamically.

Key for the exceptionally strong coupling of skyrmions to spin currents are well-understood non-vanishing Berry phases picked up when electrons or magnons traverse a skyrmion~\cite{2004_Bruno_PhysRevLett,2010_Jonietz_Science,2012_Schulz_NatPhys, 2017_Jiang_NatPhys}. Cast in the framework of an emergent electrodynamics, the skyrmion is described by virtue of an emergent magnetic field, that accounts for the deflection of conduction electrons and the concomitant appearance of a topological Hall effect of electrons traversing a skyrmion lattice~\cite{2009_Neubauer_PhysRevLett,2012_Schulz_NatPhys, 2012_Nagaosa_PhysScr, 2014_Franz_PhysRevLett}. The resulting counter-force on the skyrmion may exceed the pinning forces due to disorder such that the skyrmions start to move~\cite{2010_Jonietz_Science,2012_Schulz_NatPhys}.

Intimately connected to the spin-current-driven motion of skyrmions is the physics of their depinning~\cite{2016_Reichhardt_RepProgPhys, 2022_Reichhardt_RevModPhys} and their dynamical response~\cite{2019_Lin_PhysRevB, 2020_Seki_NatCommun, 2023_Kravchuk_PhysRevB}. For instance, an emergent electric field was observed with the onset of skyrmion motion~\cite{2012_Schulz_NatPhys, 2015_Liang_NatCommun}, and the current-induced deformation dynamics of moving skyrmion strings was tracked by means of the second-harmonic Hall effect~\cite{2018_Yokouchi_SciAdv}. Moreover, techniques such as small-angle neutron scattering~\cite{2019_Okuyama_CommunPhys, 2024_Ran_NatCommun}, ultrasound spectroscopy~\cite{2020_Luo_CommunMater}, and transmission electron or x-ray microscopy~\cite{2022_Yasin_ProcNatlAcadSciUSA, 2022_Birch_NatCommun} were combined with numerical methods to study the current-driven depinning, deformation, and creep of skyrmion lines~\cite{2016_Lin_PhysRevBa, 2017_Kagawa_NatCommun, 2020_Koshibae_SciRep, 2022_Jin_ChinPhysLett, 2023_Okumura_PhysRevLett}. In this context, open questions are whether skyrmion lines remain intact or break up into pieces by virtue of Bloch points and how they interact with surfaces~\cite{2013_Milde_Science, 2014_Schutte_PhysRevBa, 2019_Koshibae_SciRep, 2021_Birch_CommunPhys, 2022_Brearton_PhysRevB, 2023_Henderson_NatPhys, 2023_Jin_NanoLett}. 

To explore the properties of a skyrmion lattice under simultaneous application of spin transfer torques and dynamic changes of orientation, we measured the transverse susceptibility of MnSi. This compound represents the perhaps most extensively studied and best understood metallic skyrmion-hosting material. Moreover, both the unpinning of the skyrmion lattice under either electric currents and under oscillations of the magnetic field direction has been demonstrated previously in MnSi~\cite{2010_Jonietz_Science, 2016_Muhlbauer_NewJPhys}.  

As our main new result we identify, track, and interpret two key signatures: first, an increase of the real part of the susceptibility $\chi$ as the skyrmion lattice becomes unpinned and responds better to the probing field, and second, the emergence of dissipation with the onset of unpinning, encoded in a rise of the imaginary part of $\chi$, followed by a reduction of the dissipation once the skyrmions are depinned completely. We use measurements of a thermodynamic quantity, namely the magnetic AC susceptibility, in the presence of electrical currents as an unusual approach to track the depinning transition and thus the motion of skyrmions. Our experimental results are in remarkable qualitative and semi-quantitative agreement with an effective slip-stick model of the bending and motion of the skyrmions. This permits the prediction of several dynamical skyrmion lattice phases representing new phases of matter far from thermal equilibrium.

\section{Methods} 

\subsection{Sample Preparation}
A single crystal of MnSi was grown by optical float-zoning under ultra-high vacuum compatible conditions~\cite{2016_Bauer_RevSciInstrum}. X-ray Laue diffraction was used to orient the single crystal and a cuboid sample with dimensions 0.41\,mm~$\times$~0.88\,mm~$\times$~3.15\,mm was cut using a wire saw. The largest face of the sample was normal to crystallographic $ \langle001\rangle$ axis and the longest edge was parallel to a $\langle110\rangle$ axis. Current contacts were soldered to the smallest faces of the sample such that large current densities $j$ could be applied parallel to $[110]$.

\subsection{Transverse Susceptibility}
The transverse susceptibility under simultaneous application of electric currents was measured with a bespoke susceptometer in a superconducting magnet system equipped with a variable temperature Insert. The primary of the susceptometer generated an oscillatory magnetic field $\boldsymbol{b}$ perpendicular to $\boldsymbol{B}$. A concentrically aligned balanced pair of secondaries, recorded the transverse susceptibility, $\chi^{\perp}$, with respect to $\boldsymbol{B}$~\cite{2017_Rucker_PhD}. The sample equipped with electrical contacts, was mounted on a sapphire platelet inside one of the secondaries such that $\boldsymbol{b}(t)$ and $\boldsymbol{j}$ were parallel. The combination of $\boldsymbol{B}$ and $\boldsymbol{b}$ resulted in oscillations of the field direction. Data were collected for an excitation frequency of $f = 120\,\mathrm{Hz}$ and excitation amplitudes from less than $0.1\,\mathrm{mT}$ up to $12.6\,\mathrm{mT}$. The maximum amplitude was limited by the effects of ohmic heating in the current leads and the primary. In the field range of the skyrmion lattice in MnSi, between $0.16\,\mathrm{T}$ and $0.26\,\mathrm{T}$, the excitation amplitudes corresponded to small tilting angles of the total field below $5\,\mathrm{deg}$ while the field magnitude changed by less than $10^{-3}$ being essentially constant.


Further details of the experimental procedure, such as the calibration of the susceptometer, the thermal anchoring of the sample, the correction of heating effects and considerations on the field homogeneity including contributions by the Oersted field due to the currents in the sample, are reported in SI Appendix and Ref.~\onlinecite{2017_Rucker_PhD}. 

\subsection{Calculations}
We use a simple phenomenological slip-stick model, similar to Ref.~\onlinecite{2012_Schulz_NatPhys}, to describe pinning in \eqref{thieleM}. A skyrmion stays pinned, $\dot{\boldsymbol u}(z,t)=0$, when the external force at position $z$ is smaller than a critical pinning force, $|\boldsymbol F(z)|\le F^p$. In contrast, a moving skyrmion obtains a force $ \boldsymbol{ F}^{p} =-F^p (\dot{\boldsymbol{u}}/|\dot{\boldsymbol{u}|})$ oriented antiparallel to its velocity. We allow for both bulk and surface pinning, $\boldsymbol{ F}^{pin}= \boldsymbol{F}^{p,b} + \delta (z - L/2) \; \boldsymbol{F}^{p,s} + \delta (z + L/2) \; \boldsymbol{F}^{p,s}$. All numerical results are obtained by discretizing the model of \eqref{thieleM} both in time and space ($\Delta z=L/60$, $\Delta t=10^{-5}/f$ where $f=\omega/(2 \pi)$ is the excitation frequency), see SI Appendix for further details. 
	
There are effectively three fitting parameters for our theory, the dimensionless strength of bulk pinning $F^{p,b} L/(\epsilon_0)$, the ratio of surface to bulk pinning $F^{p,s}/(F^{p,s} L)$, and the characteristic frequency of the system $\omega_0=\epsilon_0/(L^2 \mathcal G)$. Furthermore, we use $\alpha \mathcal D/\mathcal{G} = 0.1$, $\beta/\alpha=0.7$. The parameter $\epsilon_0=4\cdot 10^{-13}$\,J/m is obtained from susceptibility measurements, see SI Appendix, and $v_s \approx 10^{-4}  j/j_c$ m/s, where $j_{c}\approx 10^6$ Am$^{-2}$ has been estimated in Ref.~\onlinecite{2012_Schulz_NatPhys}. This fixes all parameters of our model, including the absolute values of the susceptibilities and the values of the critical currents.

\section{Experimental Results} 

In our study we consider a static magnetic field in the $z$ direction, $B_{z}$, that is combined with a slowly oscillating field with small amplitude, $b$, in the $x$ direction, $\boldsymbol{B}(t) = (B_x(t),0,B_z)$ with $B_x(t) = b\cos(\omega t)$. We consider frequencies $\omega$ in the range of a few hertz up to a few kilohertz, which is much smaller than the gigahertz regime of typical magnons~\cite{2012_Onose_PhysRevLett, 2015_Schwarze_NatMater}. 

\begin{figure}[htb]
\centering
\includegraphics[width=1.0\linewidth]{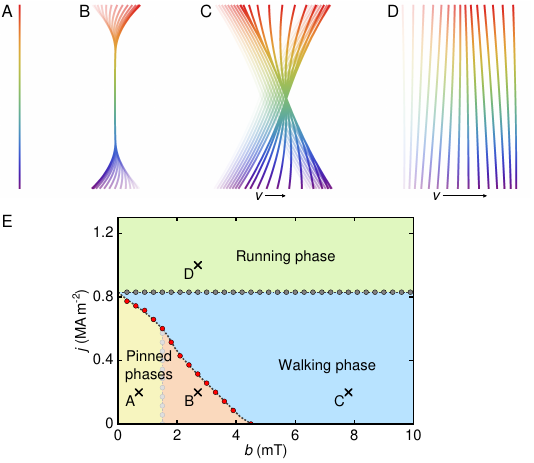}
\caption{Characteristics of skyrmion motion when an oscillating magnetic field $b$ and an electrical current density $j$ are applied perpendicular to the skyrmion lines. \mbox{({A--D})}~Visualization of the bending and motion of skyrmion lines for stroboscopic times as illustrated by decreasing color saturation. ({E})~Phase diagram of skyrmion motion obtained from simulations of \eqref{thieleM} using parameters $\epsilon_0/L F^{p,b}=36.7$, $\omega/\omega_0 = 1.26$, $F^{p,st}/(L F^{p,b})=0.2$, $\alpha \mathcal D/\mathcal{G} = 0.1$, and $\beta \mathcal D/G=0.07$ (see SI Appendix for details). Crosses denote the field and current values visualized in {A--D}. In the fully pinned (\emph{A}, yellow shading) and partially pinned phases ({B}, red shading), the skyrmion lines exhibits no net motion. In the walking ({C}, blue shading) and running phases ({D}, green shading), a net translational velocity $v$ is observed.}
\label{fig:figure1}
\end{figure}

When slowly changing the direction of the field in the absence of pinning, the skyrmion lines follow the applied field~\cite{2018_Adams_PhysRevLett}, tilting by a small angle $\alpha \approx B_x(t)/B_z$. As a consequence, the ends of the skyrmion line will move a macroscopically large distance ${\sim}L\alpha/2$, where $\alpha\ll 1$ is the tilt angle of the magnetic field and $L$ the length of the skyrmion line. 

In the presence of pinning, in contrast, whether and how the skyrmion lines follow the field direction depends on the amplitude $b$ and the frequency $\omega$ of the oscillating field. For small amplitudes, the orientation of the skyrmion line does not change. The associated magnetic texture deforms only slightly as described by standard linear-response theory. For large amplitudes, the skyrmion lines depin and follow the magnetic field, as previously reported~\cite{2016_Muhlbauer_NewJPhys}.

As depicted in Fig.~\ref{fig:figure1}, the interplay of slow oscillations of the direction of the magnetic field by a small transverse magnetic field $b$, with an applied electric current density, $j$, leads to a rich non-equilibrium phase diagram already at the level of mean-field calculations (see discussion for details). At least four different phases may be distinguished: 
(i)~In the \emph{fully pinned phase} (FPP, Fig.~\ref{fig:figure1}\,A), all parts of the skyrmion lines are static at all times with the exception of the tiny oscillations associated with magnons (not depicted in the figure).
(ii)~In the \emph{partially pinned phase} (PPP, Fig.~\ref{fig:figure1}\,B), macroscopic parts of the skyrmion lines unpin and follow the applied field, while other parts remain pinned.
(iii)~In the \emph{walking phase} (WP, Fig.~\ref{fig:figure1}\,C), all parts of a skyrmion line move over the course of a field oscillation, but remain pinned during a finite fraction of the oscillation period. We name this phase in loose analogy to the motion of humans, where walking implies that at each time at least one foot is on the ground and thus not moving.
(iv)~In the \emph{running phase} (RP, Fig.~\ref{fig:figure1}\,D), all parts of the skyrmion line are unpinned and move at all times of an oscillation.

A crucial difference of the unpinned phases as compared to the pinned phases concerns the response to an applied electric current. In the walking and the running phase arbitrarily small currents cause a net motion of the skyrmion lines despite the presence of disorder. In a similar vein, micrometer-sized skyrmion bubbles in magnetic thin films were reported to display a 300-fold increase of their diffusion constant under oscillations of a magnetic field~\cite{2023_Gruber_AdvMater}. 

Shown in Fig.\,\ref{fig:figure1}\,E is the diagram of dynamical phases as a function of oscillation amplitude and current density. At vanishingly small oscillation amplitude, the FPP of the skyrmion lattice undergoes a complete unpinning into the RP above a critical current density $j_{c}$. At zero current density, an increasing oscillation amplitude induces a transition from the FPP to the PPP, followed by the transition from the PPP to the WP. Under increasing current density, the oscillation amplitude necessary to induce the transition into the WP is reduced. Independent of the oscillation amplitude, above $j_{c}$ the WP is transformed into the RP.


\begin{figure}[htb]
\centering
\includegraphics[width=1.0\linewidth]{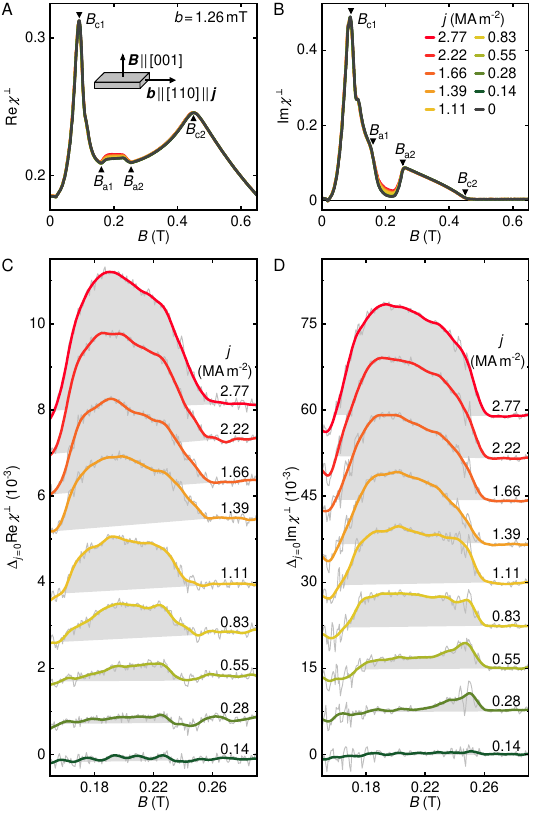}
\caption{Transverse susceptibility of MnSi as a function of static magnetic field, $B$, across the different magnetic states under various applied current densities $j$ for a medium-sized oscillating field, $b = 1.26\,\mathrm{mT}$ measured at the frequency $f=120$Hz. ({A})~Real part, $\mathrm{Re}\,\chi^{\perp}$, and ({B})~imaginary part, $\mathrm{Im}\,\chi^{\perp}$, of the transverse susceptibility. Changes of slope, labeled $B_{c1}$, $B_{a1}$, $B_{a2}$, and $B_{c2}$, indicate the transitions between the equilibrium magnetic states, see text for details. Applied currents increase the transverse susceptibility for $B_{a1} < B < B_{a2}$, i.e., in the skyrmion lattice state. \mbox{({C, D})}~Change of the real and imaginary part relative to the values observed without current, $\Delta_{j=0}\mathrm{Re}\chi^{\perp}$ and $\Delta_{j=0}\mathrm{Im}\chi^{\perp}$, in the field range of the skyrmion lattice state. Data were offset for better visibility.}
\label{fig:figure2}
\end{figure}

Before turning to our experiments, we recall that the magnetic phase diagram of MnSi includes different static magnetic phases at a temperature suitable for our study~\cite{2012_Bauer_PhysRevB}. As a function of increasing magnetic field helical order, conical order, the skyrmion lattice, conical order, and the field-polarized (ferromagnetic) state are stabilized.

For information on the experimental set-up and the precise conditions of our measurements as well as the data reduction, we refer to the Methods section and the SI Appendix. Shown in Figs.~\ref{fig:figure2}\,{A} and \ref{fig:figure2}\,{B} are the real and the imaginary parts of the transverse susceptibility, $\mathrm{Re}\chi^{\perp}$ and $\mathrm{Im}\chi^{\perp}$, as a function of the magnetic field $B$ at 28.1\,K for different current densities $j$. Starting from the helical state at $B=0$, a pronounced peak at $B_{c1}$ marks the transition to the conical state. At the lower and upper transition fields of the skyrmion lattice state at $B_{a1}$ and $B_{a2}$, changes of slope are observed. Above a kink at $B_{c2}$, the field-polarized state is entered. 

In the field range of the skyrmion lattice state, $B_{a1} < B < B_{a2}$, both $\mathrm{Re}\chi^{\perp}$ and $\mathrm{Im}\chi^{\perp}$ increase monotonically under increasing $j$. In comparison, in the helical and conical states no dependence on $j$ was observed. Relative changes of the real and the imaginary part of the susceptibility with respect to $j=0$ under various current densities are depicted in gray shading in Figs.~\ref{fig:figure2}\,{C} and \ref{fig:figure2}\,{D}. Data at $j = 0$ are consistent with previous reports of the transverse susceptibility~\cite{2015_Chacon_PhysRevLett, 2015_Nii_NatCommun}.


\begin{figure}[htb]
\centering
\includegraphics[width=1.0\linewidth]{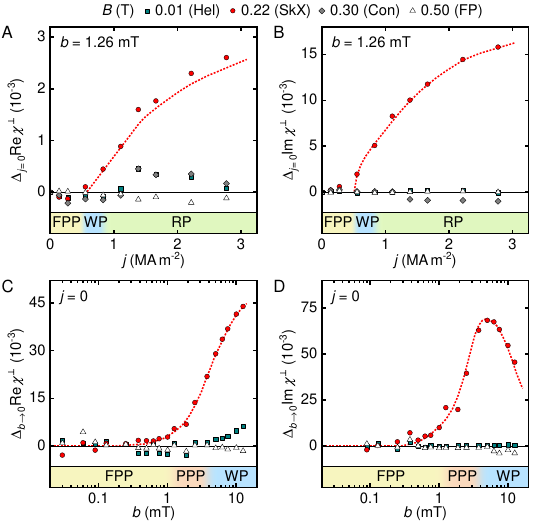}
\caption{Changes of the transverse susceptibility, $\chi^{\perp}$, in the different magnetic equilibrium states in MnSi under different applied current densities, $j$, and oscillating fields, $b$. Static field values, $B$, are representative of the helical (Hel, squares), skyrmion lattice (SkX, circles), conical (Con, diamonds), and field-polarized state (FP, triangles). Depicted in color shading below each panel are the different dynamic phases, cf. Fig.\,\ref{fig:figure1}. \mbox{({A, B})}~Change of the real and imaginary part as a function of current density relative to the values observed without current, $\Delta_{j=0}\mathrm{Re}\chi^{\perp}$ and $\Delta_{j=0}\mathrm{Im}\chi^{\perp}$. The oscillating field was $b = 1.26\,\mathrm{mT}$. \mbox{({C, D})}~Change of the real and imaginary parts as a function of the oscillation amplitude relative to the values observed at vanishing amplitude, $\Delta_{b\to0}\mathrm{Re}\chi^{\perp}$ and $\Delta_{b\to0}\mathrm{Im}\chi^{\perp}$. No current was applied, $j = 0$.}
\label{fig:figure3}
\end{figure}

We turn now to the two limiting ways of unpinning the skyrmion lattice (SkX), namely either by virtue of an electric current or oscillations of the applied field, as shown in Figs.~\ref{fig:figure3}\,{A} and \ref{fig:figure3}\,{B} or Figs.~\ref{fig:figure3}\,{C} and \ref{fig:figure3}\,{D}, respectively. For ease of comparison, additional data recorded in the helical (Hel), conical (Con), and field-polarized (FP) states are shown. 

In the skyrmion lattice phase at 0.22\,T, both $\Delta_{j=0}\mathrm{Re}\chi^{\perp}$ and $\Delta_{j=0}\mathrm{Im}\chi^{\perp}$ display essentially the same three key characteristics as a function of increasing $j$ as shown in Figs.~\ref{fig:figure3}\,{A} and \ref{fig:figure3}\,{B}. Namely, the susceptibility (i) is unchanged up to a critical current density $j_c\approx 0.5\,\mathrm{MA\,m}^{-2}$, (ii) displays an abrupt increase for $j>j_c$, followed (iii) by a sub-linear increase approaching a saturation for $j\gg j_c$. In the helical and conical states, recorded at 0.01\,T and 0.30\,T, respectively, the susceptibility remains essentially unchanged, where the small increase of $\Delta_{j=0}\mathrm{Re}\chi^{\perp}$ in the helical state (Fig.~\ref{fig:figure3}\,{A}) may be related to small systematic errors of the correction of the ohmic heating (cf. SI Appendix).

Qualitatively, $\Delta_{j=0}\mathrm{Re}\chi^{\perp}$ and $\Delta_{j=0}\mathrm{Im}\chi^{\perp}$ agree well with the behavior calculated theoretically presented below. Quantitatively the value of $j_c$ corresponds quite well to current densities in small-angle neutron scattering and the Hall effect which established that the skyrmion lattice starts to move freely~\cite{2010_Jonietz_Science, 2012_Schulz_NatPhys}. As the skyrmions are unpinned from disorder, they adjust to the small excitation field and the susceptibility increases. Thus, these characteristics in the susceptibility represent the signature of the unpinning of the skyrmion lattice under an electric current. Indicated in color shading in Fig.~\ref{fig:figure3} are the dynamical phases under increasing $j$, notably the FPP (yellow) and the RP (green). The transition must include a small regime of WP (blue) we cannot distinguish further.

Shown in Figs.~\ref{fig:figure3}\,{C} and \ref{fig:figure3}\,{D} are $\Delta_{j=0}\mathrm{Re}\chi^{\perp}$ and $\Delta_{j=0}\mathrm{Im}\chi^{\perp}$ as a function of oscillation amplitude, ${b}$, at $j=0$. The key characteristics observed as a function of $b$ are similar though not identical to those observed under increasing $j$. Namely, $\Delta_{j=0}\mathrm{Re}\chi^{\perp}$ and $\Delta_{j=0}\mathrm{Im}\chi^{\perp}$ are unchanged up to a critical amplitude $b_c\approx 0.7\,\mathrm{mT}$. They initially exhibit a monotonic increase for $b>b_c$. However, $\Delta_{j=0}\mathrm{Re}\chi^{\perp}$ and $\Delta_{j=0}\mathrm{Im}\chi^{\perp}$ dsiplay a point of inflection and a maximum, respectively, around $b^*\approx 5\,\mathrm{mT}$ beyond which they do not track each other. In comparison to the skyrmion lattice phase (SkX), data recorded in the helical phase (Hel) at 0.01\,T and conical phase (Con) at 0.30\,T are essentially constant as a function of $b$.

The dependence on oscillation amplitude is also in remarkable qualitative agreement with the theoretical calculation presented below. The unpinning permits the skyrmions to align better with the applied field, resulting in an increase of the real part of the susceptibility. The maximum observed in $\Delta_{j=0}\mathrm{Im}\chi^{\perp}$ reflects dissipation which decreases as the skyrmions become fully unpinned. The color shading denotes the field range of the FPP (yellow), the PPP (orange), and the WP (blue). As $j=0$, the skyrmions are walking on the spot in the WP. Underscoring this interpretation quantitatively, the value of $b_c\approx 0.7\,\mathrm{mT}$ translates to a tilting angle of ${\sim}0.2^{\circ}$ compared to a tilting angle for unpinning ${\sim}0.4^{\circ}$ observed microscopically in TISANE~\cite{2016_Muhlbauer_NewJPhys}. 


\begin{figure}[htb]
\centering
\includegraphics[width=1.0\linewidth]{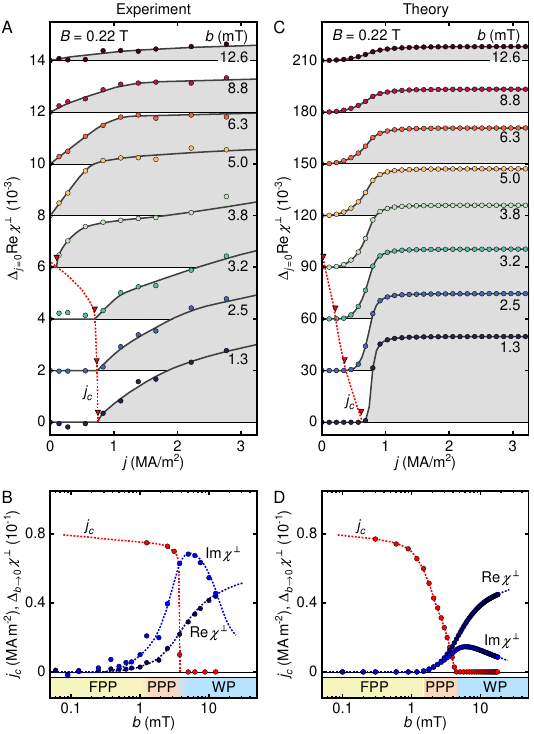}
\caption{Suppression of the critical current density, $j_{c}$, of skyrmion lattice flow under transverse oscillating field, $b$. Experimental results (left column) are compared with theoretical calculations (right column). The skyrmion lattice state was stabilized at the static magnetic field $B = 0.22\,\mathrm{T}$. \mbox{(\emph{A, C})}~Change of the real part of the transverse susceptibility as a function of current density $j$ relative to the values observed without current, $\Delta_{j=0}\mathrm{Re}\chi^{\perp}$, for different oscillation amplitudes, $b$. \mbox{(\emph{B, D})}~Change of the real and imaginary part of the transverse susceptibility (blue symbols) as function of the oscillation amplitude relative to the values observed at vanishing amplitude, $\Delta_{b\to0}\mathrm{Re}\chi^{\perp}$ and $\Delta_{b\to0}\mathrm{Im}\chi^{\perp}$, when no current is applied, $j = 0$. When current and field oscillations are combined, the critical current density $j_{c}$ of net translational motion of the skyrmion lines (red symbols) is suppressed to zero under increasing amplitude $b$, i.e., the walking phase is observed already at arbitrarily small current density. For the calculations, parameters were used as specified in the caption of Fig.~\ref{fig:figure1}.}
\label{fig:figure4}
\end{figure}

The main result of our study is finally obtained when combining different current densities with different field oscillation amplitudes. Shown in Fig.~\ref{fig:figure4}\,{A} is $\Delta_{j=0}\mathrm{Re}\chi^{\perp}$ as a function of current density for different oscillation amplitudes $b$. At small amplitudes the critical current density $j_{c}$, defined as the point where $\Delta_{j=0}\mathrm{Re}\chi^{\perp}$ increases abruptly (red triangles), decreases and vanishes with increasing $b$. Shown in Fig.~\ref{fig:figure4}\,{B} is a comparison of $j_c$ with the change of the susceptibility of the skyrmion lattice induced by $b$ at $j=0$, as already presented in Figs.~\ref{fig:figure3}\,{C} and \ref{fig:figure3}\,{D}. For $b>b^*$ the critical current density $j_c$ vanishes, i.e., the amplitude at which the skyrmion lattice becomes completely unpinned is at $j=0$. Thus, under strong oscillations of the field direction a dynamical phase of the skyrmion lattice is created, in which arbitrarily small current densities $j$ may generate a net motion. 


\section{Discussion} 


For a minimal theoretical description of our experimental observations, we developed a theory which includes, (i),~non-linear effects due to the coupling of a skyrmion line to a magnetic field of oscillating direction, (ii),~the forces on the skyrmion lattice due to spin-polarized electric currents, and, (iii),~an effective account of pinning by disorder. The effects of magneto-crystalline anisotropies are neglected being very weak in MnSi~\cite{2017_Bauer_PhysRevB, 2018_Adams_PhysRevLett}.

We consider a small time-dependent transverse magnetic field component, $b \cos(\omega t)$, that induces a tilt of the skyrmion lattice assuming that the skyrmion lattice is free to move. To describe the tilting and any associated bending as well as a motion of the skyrmion lattice, we introduce a function $\boldsymbol{ u}(z,t)=(u_x(z,t),u_y(z,t),z)$ which describes the displacement of skyrmion lines of length $L$, where $z$ with $\frac{-L}{2}\le z \le \frac{L}{2}$ is a coordinate parallel to the skyrmion lines.

Assuming that the magnetization can be described by 
$ \boldsymbol{ M}(\boldsymbol{ r},t) \approx \boldsymbol{ M}_0(\boldsymbol{ r}-\boldsymbol{ u}(z,t))$ 
where $\boldsymbol{M}_0$ is the skyrmion lattice in equilibrium, we derive the Thiele equation for $u(z,t)$:
\begin{equation}\label{thieleM}
	\mathcal{G} \times (\dot{\boldsymbol{u}}(z,t) -\boldsymbol{v_s}) + \mathcal{D} ( \alpha\dot{\boldsymbol{u}}(z,t) -   \beta \boldsymbol{v_s}) = -\frac{\delta \mathcal F_u}{\delta \boldsymbol{ u}(z,t)} + \boldsymbol{ F}^{\mathrm{pin}}.
\end{equation}
Here, $\mathcal{G}$ is the gyrocoupling vector, $\mathcal{D}$ is the dissipative tensor,  $\boldsymbol{v_s}$ is a (spin) drift velocity proportional to the applied electric current, $\alpha$ and $\beta$ parameterize the Gilbert damping, and $\boldsymbol{ F}^{\mathrm{pin}}$ is an effective pinning force due to disorder. 
The free energy describing the alignment of the skyrmion lines parallel to the applied magnetic field is given by
\begin{align}\label{FF}
	\mathcal F_u \approx \frac{\epsilon_0}{2}\int_{-L/2}^{L/2} dz \;  \left( \partial_z \boldsymbol{u}- \frac{\boldsymbol{B}_x(t)}{B} \right)^2
\end{align}
where $\boldsymbol{B}_x(t) = b\cos(\omega t)\hat{\boldsymbol{x}}$ and $\epsilon_0$ is is an elastic constant parametrizing the stiffness of the skyrmion lattice. We refer to SI Appendix for details and derivations. 

The effect of disorder is typically not included in Thiele-style equations for skyrmion dynamics, but proves to be essential in our study. For an effective description we use a slip--stick model, where the pinning forces at the surface and in the bulk differ, similar to models previously applied to skyrmions~\cite{2012_Schulz_NatPhys} and superconducting vortices~\cite{2018_Liarte_PhysRevAppl}, see Methods and SI Appendix. 

\eqref{thieleM} represents an effective mean-field description of the bending and the translational motion of a skyrmion lattice, where fluctuation effects are ignored and disorder is only taken into account on average. Thus, this model captures the depinning transition only on a mean-field level and is not able to describe the behavior in regimes in which the physics of thermal creep is important. Yet, as a mean-field theory, it captures the relevant phases and phase transitions.

As expressed in \eqref{FF}, the transverse magnetic field, $b$, couples to the end of the skyrmion string only, 
$\int_{-L/2}^{L/2}\partial_z \boldsymbol u dz=\boldsymbol{ u}(L/2) - \boldsymbol{u}(-L/2)$. Thus, the transverse magnetic field enters only on the level of a boundary condition of \eqref{thieleM}. Shown in Fig.~\ref{fig:figure1} is the phase diagram when the pinning at both surfaces is identical but differs from the bulk. As presented in the introduction, one may then distinguish four types of phases, namely fully pinned (FPP), partially pinned (PPP), walking (WP), and running (RP). When the pinning at the surfaces differs in addition, the partially pinned phase splits up into several sub-phases (see SI Appendix for details). 


The key question pursued in our study concerned the signatures of the different dynamical phases and associated phase transitions in $\mathrm{Re}\chi^{\perp}$ and $\mathrm{Im}\chi^{\perp}$ as a function of oscillation amplitude $b$ and electrical current density $j$. Shown in Figs.~\ref{fig:figure4}\,C and \ref{fig:figure4}\,D are changes of the calculated transverse susceptibility, $\Delta_{b\to0}\mathrm{Re}\chi^{\perp}$ and $\Delta_{b\to0}\mathrm{Im}\chi^{\perp}$, for a set of parameters chosen such that the values of the critical current density $j_{c}$ at $b=0$ and the real part of the susceptibility, $\mathrm{Re}\chi^{\perp}$ at zero current match experiment.

As a function of current density, the real part of the calculated susceptibility, $\Delta_{b\to0}\mathrm{Re}\chi^{\perp}$, shown in Fig.~\ref{fig:figure4}\,C, displays essentially all of the key characteristics observed experimentally. Namely, we observe (i), no change up to $j_c$, (ii), a pronounced increase at $j_c$, and, (iii), sublinear behavior approaching saturation for $j\gg j_c$. The same agreement is also seen in the imaginary part, $\mathrm{Im}\chi^{\perp}$ (cf. SI Appendix). 

This evolution may be understood as follows. A constant current in the PPP causes a static deformation even in the presence of an oscillating magnetic field. Therefore, the difference $\Delta_{j=0}\chi^{\perp}(b,j) = \chi^{\perp}(b,j) - \chi^{\perp}(b,j=0)$, in both the real and imaginary parts vanishes exactly within the PPP. Only when entering the WP or RP, the response measured by the real or imaginary part of $\Delta_{j=0} \chi^{\perp}(b,j)$ is affected by the current. Thus, the current response of $\Delta_{j=0} \chi^{\perp}$, shown for $\mathrm{Re}\chi^{\perp}$ in Fig.~\ref{fig:figure4}\,C, permits to identify the transition into the WP. 

Shown in Fig.~\ref{fig:figure4}\,D is the calculated amplitude dependence of $\mathrm{Re}\chi^{\perp}$ and $\mathrm{Im}\chi^{\perp}$ for $j=0$ and of $j_c$ as inferred from the current dependence shown in Fig.~\ref{fig:figure4}\,C. Again excellent qualitative agreement with experiment is observed. For small amplitudes, $b < 0.7\,\mathrm{mT}$, the skyrmion lattice is in the fully pinned phase (FPP). Under increasing amplitude, both $\mathrm{Re}\chi^{\perp}$ and $\mathrm{Im}\chi^{\perp}$, start to increase at the transition from the fully pinned phase (FPP) to the partially pinned phase (PPP). 

Further, the transition between the PPP and the WP for $j=0$ does not exhibit a specific signature in the susceptibility (there is no net velocity in either phase). Instead a point of inflection in $\mathrm{Re}\chi^{\perp}$ that coincides with the maximum in $\mathrm{Im}\chi^{\perp}$ is predicted. Such a maximum in $\mathrm{Im}\chi^{\perp}$ is indeed observed experimentally as shown in Fig.~\ref{fig:figure4}\,B. The decrease of $\mathrm{Im}\chi^{\perp}$ and hence dissipation following the maximum represents evidence of the complete unpinning of the skyrmion lattice. 


While our mean-field theory is in remarkable agreement with experiment overall, we observe also several important discrepancies. The increase of $\mathrm{Re}\chi^{\perp}$  as a function of current density observed experimentally is much smaller than in our theory (Figs.~\ref{fig:figure4}\,A and \ref{fig:figure4}\,C). Furthermore, in the walking phase, i.e., when $j_{c}$ is suppressed to 0 at large amplitudes, theory predicts an analytic response to the current, $\Delta_{j=0}\mathrm{Re}\chi^{\perp} \propto j^2$ for small $j$, while experiment suggests $\Delta_{j=0}\mathrm{Re}\chi^{\perp} \propto |j|^\alpha$ with $\alpha \approx 1$.

A quantitative discrepancy is also visible in  $\mathrm{Im}\chi^{\perp}$ as a function of oscillation amplitude. While qualitatively the same field-dependence is observed in our theory, the absolute value of dissipation  is approximately a factor two larger  in the experiment. Finally, values of $j_c$ observed experimentally at intermediate values of $b$ are larger than those calculated theoretically (Figs.~\ref{fig:figure4}\,B and \ref{fig:figure4}\,D).

A large part of the observed discrepancies may arise from the mean-field nature of our theory. The theory does not capture that only a fraction of skyrmions may move in the presence of currents and that moving skyrmion lattices becomes less disordered~\cite{2016_Reichhardt_RepProgPhys, 2022_Reichhardt_RevModPhys}. This may explain the non-analytic onset of $\Delta_{j=0}\mathrm{Re}\chi^{\perp} $ and the reduced amplitude of this quantity seen experimentally. Furthermore, an asymmetry of surface pinning potentials may explain an increased amplitude of $\mathrm{Im}\chi^{\perp}$ and a more abrupt drop of $j_c$ as function of $b$ (see SI Appendix). Another potentially important effect is that skyrmion lines may break up into pieces as investigated numerically in Ref.~\onlinecite{2019_Koshibae_SciRep}. The bending of skyrmion lines by the oscillating field, Fig.~\ref{fig:figure1}, and the associated elastic forces may favor such a breaking-up (see SI Appendix 2H). As smaller skyrmion pieces can follow the magnetic field more easily, this leads to larger values of $\chi^\perp$ as a function of $b$ as compared to the increase as a function of $j$.

\section{Conclusions} 

When manipulating magnetic textures for the creation of novel non-equilibrium states of matter \cite{Romain2024,Hardt2025} pinning by disorder is arguably one of the most important constraints. Access and control of pinning processes are of central importance for all skyrmion-based non-equilibrium properties as well as the creation of technologically useful skyrmion devices. In our study, we showed that periodically changing the magnetic field direction by just a few degrees provides access to dynamic phases in which the pinning is effectively overcome. When combined with (spin) currents a phase, in which skyrmions move at arbitrarily small current densities, becomes part of a rich phase diagram. The non-equilibrium character of these phases permits to resolve a wide range of fundamental and applied questions in future studies. 


\begin{acknowledgments}
We wish to thank S.\ Mayr for fruitful discussions and assistance with the experiments. This study has been funded by the Deutsche Forschungsgemeinschaft (DFG, German Research Foundation) under CRC 1238 (Control and Dynamics of Quantum Materials, Project No. 277146847, subproject C04), CRC/TR 183 (Entangled States of Matter, Project No. 277101999, subproject A01), TRR80 (From Electronic Correlations to Functionality, Project No.\ 107745057), TRR360 (Constrained Quantum Matter, Project No. 492547816), SPP2137 (Skyrmionics, Project No.\ 403191981, Grant PF393/19), and the excellence cluster MCQST under Germany's Excellence Strategy EXC-2111 (Project No.\ 390814868). Financial support by the European Research Council (ERC) through Advanced Grants No.\ 291079 (TOPFIT) and No.\ 788031 (ExQuiSid) is gratefully acknowledged.
\end{acknowledgments}


\end{document}


\setcounter{secnumdepth}{3}
\setcounter{equation}{0}
\setcounter{figure}{0}
\renewcommand{\theequation}{S-\arabic{equation}}
\renewcommand{\thefigure}{S\arabic{figure}}
\renewcommand{\thetable}{S\Roman{table}}
\newcommand\Scite[1]{[S\citealp{#1}]}




 \clearpage
\newpage




%
%
%


\title{Supplementary Information for\\
"Shaking and pushing skyrmions: Formation of a non-equilibrium phase with zero critical current"}

\author{Felix Rucker}
 \affiliation{School of Natural Sciences, Technical University of Munich, D-85748 Garching, Germany}

\author{Alla Bezvershenko}
 \affiliation{Institute for Theoretical Physics, Universit\"at zu K\"oln, D-50937 K\"oln, Germany}

\author{Denis Mettus}
 \affiliation{School of Natural Sciences, Technical University of Munich, D-85748 Garching, Germany}
 
\author{Andreas Bauer}
 \affiliation{School of Natural Sciences, Technical University of Munich, D-85748 Garching, Germany}
 \affiliation{Zentrum f\"ur QuantumEngineering (ZQE), Technical University of Munich, D-85748 Garching, Germany}
 
\author{Markus Garst}
\affiliation{Institut f\"ur Theoretische Festk\"orperphysik, Karlsruhe Institute of Technology, D-76131 Karlsruhe, Germany}
\affiliation{Institute for Quantum Materials and Technology, Karlsruhe Institute of Technology, D-76131 Karlsruhe, Germany}
 
\author{Achim Rosch}
 \affiliation{Institute for Theoretical Physics, Universit\"at zu K\"oln, D-50937 K\"oln, Germany}

\author{Christian Pfleiderer}
\affiliation{School of Natural Sciences, Technical University of Munich, D-85748 Garching, Germany}
\affiliation{Zentrum f\"ur QuantumEngineering (ZQE), Technical University of Munich, D-85748 Garching, Germany}
\affiliation{Munich Center for Quantum Science and Technology (MCQST), Technical University of Munich, D-85748 Garching, Germany} 
\affiliation{Munich Center for Quantum Science and Technology (MCQST), Technical University of Munich, D-85748 Garching, Germany} 
\affiliation{Heinz Maier-Leibnitz Zentrum (MLZ), Technical University of Munich, D-85748 Garching, Germany}

\date{\today}

\begin{abstract}
In this supplemental information (SI), we present details of the experimental methods, and the effective model used to describe skyrmions under shaking and pushing.
\end{abstract}

\maketitle

\section{SI Methods}
\label{SI methods}
\subsection{Sample}
\begin{figure*}[htb!]
	\centering
	\includegraphics[width=1\linewidth]{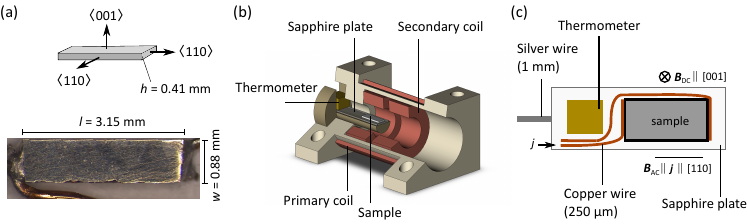}
	\caption{Depictions of sample, susceptometer, and sample holder. (a)~Sketch of the  shape and orientation of the sample; photograph of the MnSi sample with the current leads attached. (b)~Schematic cut-away view of the bespoke transverse AC susceptometer comprising a primary with a balanced pair of secondaries. (c)~Schematic depiction of the experimental arrangement comprising the sample and a thermometer as mounted on a sapphire platelet that was anchored thermally using a silver wire. Electric currents (density $j$) and the AC excitation field, denoted $B_{AC}$, were applied in parallel along a crystallographic $\langle110\rangle$ direction. The static field, denoted $B_{DC}$, was oriented perpendicular to $B_{AC}$, along a $\langle100\rangle$ direction. The thermometer was attached close to the sample to ensure accurate temperature readings.}
	\label{fig:figureS01}
\end{figure*}

Single-crystal MnSi was prepared by optical float zoning following the procedure reported in Ref.\,\cite{2010_Bauer_PhysRevB}. X-ray Laue diffraction was used to orient the single crystal. A cuboid sample was cut using a wire saw. The sample dimensions were 0.41\,mm~$\times$~0.88\,mm~$\times$~3.15\,mm with a crystallographic $\langle001\rangle$ axis normal to the largest surface and a $\langle110\rangle$ axis parallel to the longest side of the sample. Current contacts were soldered to the smallest sides of the sample for electric currents along the crystallographic $\langle110\rangle$ direction. The sample orientation and a photograph of the sample including the current contacts are shown in Fig.\,\ref{fig:figureS01}\,(a). The sample had a residual resistivity ratio of 80, comparable to typical samples studied in the literature~\cite{2010_Bauer_PhysRevB}.

\subsection{Susceptometer} 
For our study a bespoke AC susceptometer was developed as shown schematically in Fig.\,\ref{fig:figureS01}\,(b)~\cite{2017_Rucker_PhD}. The susceptometer was mounted on a PEEK (polyether ether ketone) body which served also as the coil former of the primary. The primary was wound directly onto the PEEK support using a 100~$\mu$m Rutherm VB 155 enamelled copper wire. A set of balanced secondaries was attached to the inside of the PEEK support. The sample was attached to a PEEK sample holder using GE varnish. To track the sample temperature a sapphire platelet was attached to the sample holder supporting the sample as well as a thermometer both shown schematically in Fig.\,\ref{fig:figureS01}\,(c). The sapphire platelet was anchored thermally by means of a silver wire (diameter 1\,mm). The voltage induced in the balanced pair of secondaries was calibrated against the susceptibility recorded with the ACMS II option of a Quantum Design Physical Properties Measurement System~(PPMS) in the same MnSi sample.


\subsection{Correction of ohmic heating by the sample}
A key aspect of the evaluation of the experimental data concerned the determination of the true sample temperature. Data were recorded as a function of magnetic field, where the temperature of the He gas stream was carefully regulated to be constant and the temperature of the sapphire platelet recorded. In the following the procedur used to estimate and correct for the ohmic heating by the sample is described.

\begin{figure}[htb!]
	\centering
	\includegraphics[width=1\linewidth]{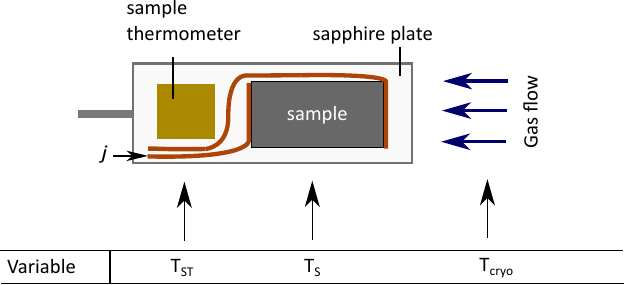}
	\caption{Schematic depiction of the temperatures as defined in our experiments. $T_\text{ST}$ represents the temperature detected by the sample thermometer attached to the sapphire platelet. $T_\text{s}$ denotes the true sample temperature. Due to ohmic heating in the sample a difference exists between $T_\text{s}$ and $T_\text{ST}$, denoted $\Delta T_\text{sample} = T_\text{s} - T_\text{ST}$. $T_\text{cryo}$ represents the control temperature of the gas stream in the sample chamber of the cryostat.}
	\label{fig:figureS02}
\end{figure}

Shown schematically in Fig.\,\ref{fig:figureS02} is the experimental setup and the associated characteristic temperatures. The control temperature of the cryostat, $T_\text{cryo}$, represents the temperature of the gas flow in the sample chamber of the VTI. The temperature of the sapphire platelet, $T_\text{ST}$, supporting the sample was recorded with a calibrated \textit{Lakeshore Cernox} sensor, mounted close to the sample. The DC current applied to the sample generated ohmic heating in the sample. This caused a temperature difference between the true sample temperature, $T_\text{s}$, and the temperature recorded with the sample thermometer, $T_\text{ST}$. This difference depends on the following parameters: 
\begin{compactitem}
	\item[(a)] the cooling power $P$ of the cryostat, which varies as a function of temperature;
	\item[(b)] the applied current $I$ in the sample causing ohmic heating;
	\item[(c)] the resistivity $\rho_\text{xx}$ of the sample, which changes as a function of temperature $T$ and the applied magnetic field $B$.
\end{compactitem}

\begin{figure}[htb!]
	\centering
	\includegraphics[width=1\linewidth]{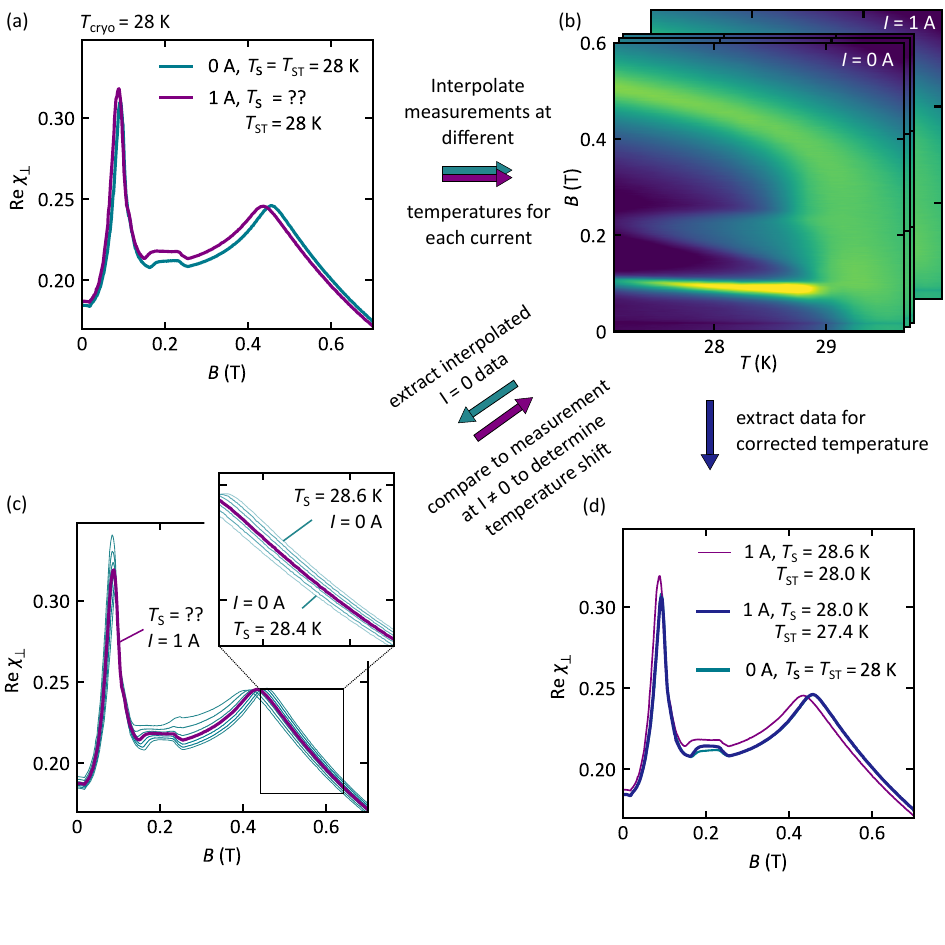}
	\caption{Flow chart of the procedure used to account for systematic temperature differences due to ohmic heating between the true sample temperature and the thermometer. (a) Typical data as recorded at $T_\text{cryo}= 28$\,K and DC currents $I=0$ and $I=1$\,A. Due to ohmic heating, the temperature of the measurements at non-zero current is unknown. (b) Interpolating of magnetic field sweeps at different temperatures for each DC current 2D data grids are generated . (c) By minimizing the difference between the high current measurement (purple curve) and the zero current data grid (light green curves) in the field polarized region between 450\,mT and 650\,mT, the effective sample temperature can be determined, assuming a current independent susceptibility in this region. (d) The determined sample temperature is used to extract the high current measurement curve from the corresponding grid which can then be compared to measurements at the same sample temperature, but different DC current (blue and green curve).}
	\label{fig:figureS03}
\end{figure}

Summarized in Fig.\,\ref{fig:figureS03} is the procedure used to infer $\Delta T_\text{sample}$, when neglecting the magnetic field dependence of the sample resistance. We return to the influence of the magnetic field dependence of the sample resistance, which proves to be negligibly small, in the next section. Shown in Fig.\,\ref{fig:figureS03}\,(a) are typical susceptibility data recorded at the same temperature of the sample thermometer, $T_\text{ST} = 28$\,K, under DC currents $I=0$ and $I=1$\,A applied to the sample. Differences of the data are in parts due to a deviation of the true sample temperature from the measured sample temperature, $\Delta T_\text{sample} = T_\text{s}-T_\text{ST}$. 

For the determination of the true sample temperature when neglecting the magnetic field dependence of the sample resistance, we assume that the transverse susceptibility in the field-polarized phase of MnSi is unchanged as a function of DC current $I$. This assumption appears plausible, as spin torque effects in the ferromagnetic regime are only expected at current densities exceeding the values applied in our study by many orders of magnitude. In turn, the transverse susceptibility $\chi_\perp$ was recorded as a function of the applied magnetic field between $B=0$ and $B=0.7$\,T at selected temperatures between $T=27$\,K and $T=29$\,K, notably the temperature range of the skyrmion lattice phase. These measurements were systematically repeated at different applied currents between $I=0$ and $I=1$\,A (cf. Fig.\,\ref{fig:figureS03}\,(b)). All data recorded at the same DC current were interpolated on a fine two-dimensional grid to infer the transverse susceptibility $\chi_\perp$ as a function of magnetic field $B$. 

Following this, each data set recorded at non-zero current was compared to the two-dimensional grid determined in magnetic field sweeps at zero current. Shown in Fig.\,\ref{fig:figureS03}~(c) are typical data at non-zero current (purple) and a set of magnetic field sweeps at different temperatures, extracted from the 2D grid for $I=0$ (light green). Assuming purely ohmic heating in the field-polarized state above $B\approx450$\,mT (i.e., the absence of spin torque contributions to $\chi_\perp$), the susceptibility observed under magnetic field sweeps recorded at zero DC current and finite DC current are expected to match between $B=0.45$\,T and $B=0.65$\,T as the sample is at the same temperature. In turn, the data at zero current matching best with the data at non-zero current in the field-polarized state were identified. As $T_\text{s}$ is known for the data recorded at zero current, we assume that the data recorded at non-zero current was recorded at the same sample temperature. As a final step, the temperature of the two-dimensional grids of data recorded at non-zero current was corrected. The magnetic field dependence was then inferred for different applied electrical currents from the temperature corrected grids, cf. Fig.\,\ref{fig:figureS03}\,(d). 

\begin{figure}[htb!]
	\centering
	\includegraphics[width=0.8\linewidth]{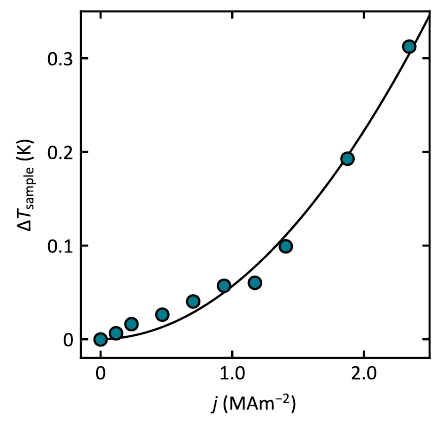}
	\caption{Temperature shift, $\Delta T_\text{sample}$, of the sample temperature $T_\text{S}$ compared to the temperature recorded by the sample thermometer, $T_\text{ST}$, as a function of applied DC current density $j$. Data points represent the temperature shift determined by the procedure shown in Fig.\,\ref{fig:figureS03}. The black line shows a quadratic fit consistent with ohmic heating.}
	\label{fig:figureS04}
\end{figure}

Shown in Fig.\,\ref{fig:figureS04} is the average temperature difference $\Delta T_\text{sample}$ determined this way. $\Delta T_\text{sample}$ varies quadratically as a function of the applied current consistent with ohmic heating and a constant cooling power of the system. This temperature correction was determined for all susceptibility data recorded at non-zero current.


\subsection{Role of the magnetic field dependence of the sample resistance}

\begin{figure*}[htb!]
	\centering
	\includegraphics[width=0.8\linewidth]{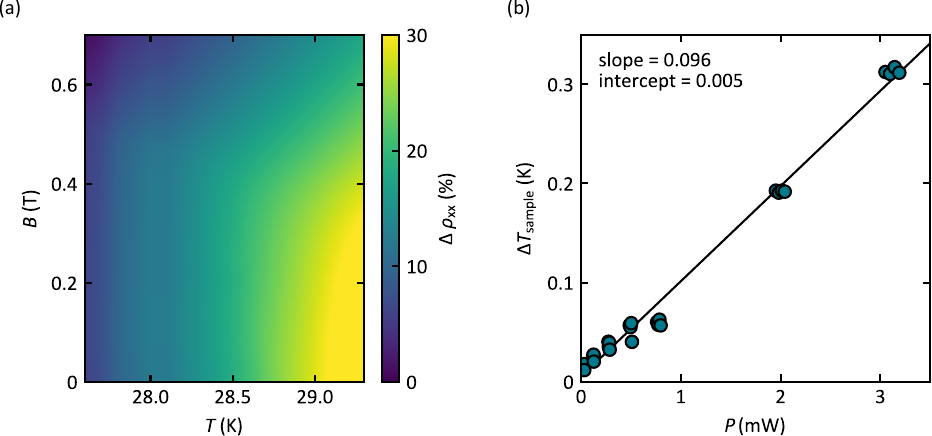}
	\caption{Connection between $\Delta T_\text{sample}$ and the ohmic power deposited in the sample, $P$. (a) Variation of the resistivity as a function of sample temperature $T$ and applied magnetic field $B$. Data reported in Ref.\,~\cite{2014_Franz_PhysRevLett} were used to calculate this variation. (b) $\Delta~T_{\text{sample}}$ as a function of the heat generated by the DC current applied. To calculate the heating power, the sample resistance was inferred from panel (a). The parameters of the linear fit may be used to quantify the correction of the change of sample temperature due to changes of resistance under magnetic field.}
	\label{fig:figureS05}
\end{figure*}

\begin{figure*}[htb!]
	\centering
	\includegraphics[width=0.7\linewidth]{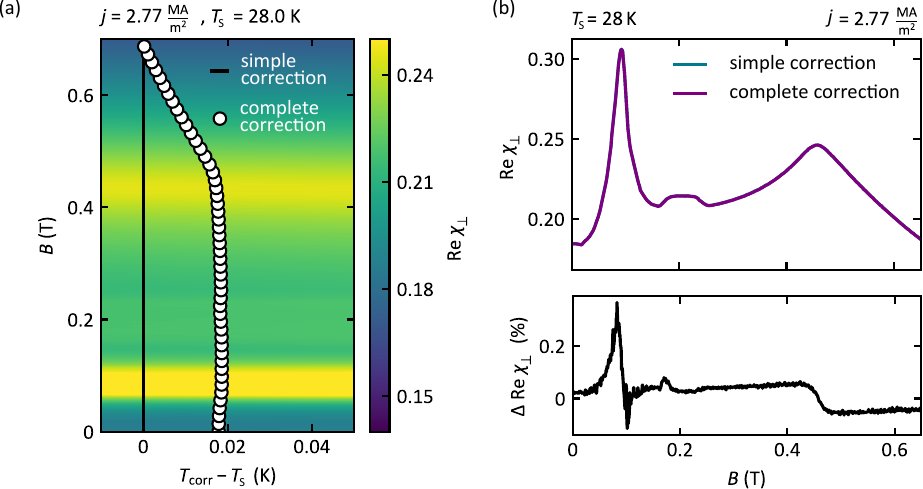}
	\caption{Estimate of changes of the true sample temperature due to the magnetic field dependence of the sample resistance. (a)~Two-dimensional interpolation of the real part of the susceptibility, $\mathrm{Re}\chi_\perp$ as a function of temperature $T$ and magnetic field $B$ in the vicinity of a sample temperature $T_\text{s}=28.0\,{\rm K}$. Values on the horizontal axis are denoted relative to $T_\text{s}$ (for details see text). (b)~Susceptibility inferred from panal (a) along the black line (simple correction, turquoise curve) neglecting the field dependence of the sample resistance, and the white dots taking into account the magnetic field dependence of the sample resistance (complete correction, purple line). (c)~Difference of $\mathrm{Re}\chi_\perp$ inferred with and without magnetic field dependence of the sample resistance. The contribution by the magnetic field dependence of the sample resistance is negligible.}
	\label{fig:figureS06}
\end{figure*}

Figs.\,~\ref{fig:figureS05} and \ref{fig:figureS06} summarize the correction of changes of ohmic heating due to the magnetic field dependence of the sample resistance. To estimate the heat deposited in the sample, data reported in Ref.\,\cite{2014_Franz_PhysRevLett} were used to compute variations of the resistivity as a function of temperature and magnetic field shown in Fig.\,\ref{fig:figureS05}\,(a). To estimate the shift of sample temperature as a function of the ohmic heating power $P$ generated in the sample, the temperature correction described above was performed for selected magnetic field sweeps. 

As the resistivity $\rho_{xx}(T,B)$ is known as a function of magnetic field and temperature, $\Delta T_\text{sample}$ for each curve may be connected to the heating power in terms of a linear interpolation. This interpolation is required, as the cooling power of the system is unknown. $\Delta T_\text{sample}$ of selected measurements at various heating power are shown in Fig.\,\ref{fig:figureS05}\,(b). Evaluating all data, the following relation was obtained for our setup:
\begin{equation}
	\Delta T_\text{sample} (P)  = 0.096\,\frac{\mathrm{K}}{\mathrm{mW}} \cdot P + 0.005\,\mathrm{K}.
\end{equation}
Here, $P$ is a function of the applied current $I$ and the sample resistance $R_{xx}(B)$:
\begin{equation}
	P (I, \rho_{xx}(B)) = I^2\cdot R_{xx}(B).
\end{equation}

Next, we estimated the corrected true sample temperature $T_\text{corr}(B)$, when taking into account the magnetic field dependence of the sample resistance. Fig.\,\ref{fig:figureS06}\,(a) shows a two-dimensional interpolation of $\mathrm{Re}\chi_\perp$ as a function of magnetic field and temperature. The temperature axis is denoted with respect to the sample temperature $T_s$ as determined above in units $T_\text{corr}-T_\text{s}$ (for the data shown here $T_s=28.0\,{\rm K}$). 

The black vertical line indicates the magnetic field dependence along which the susceptibility values were extracted when neglecting the magnetic field dependence of the sample resistance. In comparison, the white dots shown in Fig.\,\ref{fig:figureS06}\,(a) indicate the trajectory of the magnetic field dependence when taking into account the magnetic field dependence of the sample resistance. The additional temperature correction is expected to be of the order of 20\,mK for the largest DC currents investigated. 

Shown in Fig.\,\ref{fig:figureS06}\,(b) is the real part of the susceptibility, where the turquoise curve neglects the magnetic field dependence of the sample resistance (black line in Fig.\,\ref{fig:figureS06}\,(a)), while the purple curve takes into account the magnetic field dependence of the sample resistance (white dots in Fig.\,\ref{fig:figureS06}\,(a)). No significant differences between the two data sets are visible. Shown in Fig.\,\ref{fig:figureS06}\,(c) is the relative deviation of the data sets shown in Fig.\,\ref{fig:figureS06}\,(b). For the largest currents applied, the difference between both data sets varies between 0.1\% and 0.3 \% and is thus negligible. 

\subsection{Oersted magnetic fields}
\begin{figure*}[htb!]
	\centering
	\includegraphics[width=0.8\linewidth]{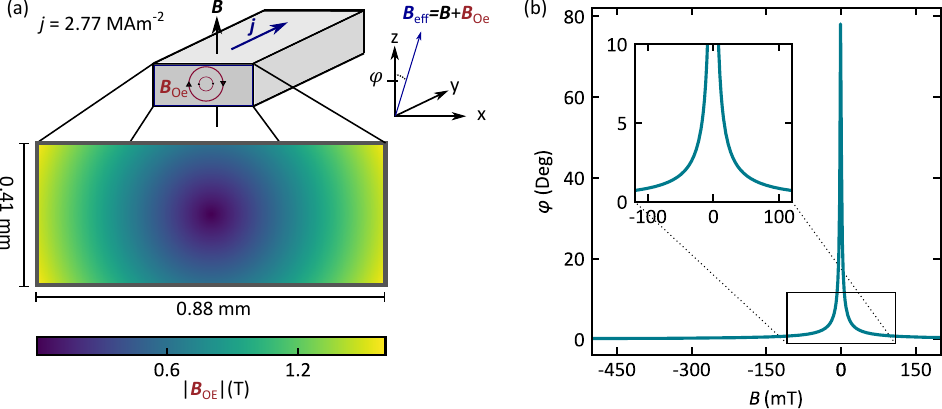}
	\caption{Effects of Oersted fields, $\boldsymbol{B}_\text{Oe}$, generated by the DC current flow $j$ in the sample. \mbox{(a) Cross-sectional} depiction of the sample at the maximum current of  \mbox{$j = 2.77 \frac{\mathrm{MA}}{\mathrm{m}^2}$}. Oersted fields of up to $\boldsymbol{B}_\text{Oe} = 1.6$\,mT are generated at the edges of the sample. (b) Resulting effective magnetic field angle $\varphi$ with respect to the DC field axis as a function of the applied field, $\boldsymbol{B}$. Inset shows a close up view for small applied fields.}
	\label{fig:figureS07}
\end{figure*}

As a result of the applied DC current, Oersted fields arise within the sample and change the magnetic field strength and distribution of magnetic field values within the sample. Fig.\,\ref{fig:figureS07}\,(a) shows a schematic drawing of the experimental set up, indicating the directions of current flow $\boldsymbol{j}$ and magnetic field $\boldsymbol{B}$ as well as the Oersted fields (red circles) in the sample. For current along the $y$-direction, the Oersted fields are generated in the $x$-$z$-plane.

A cross-sectional view of the absolute value of the Oersted fields generated in the sample,  $\left|\boldsymbol{B}_\text{Oe}\right|$, is shown in Fig.\,\ref{fig:figureS07}\,(a). The largest field value arise at the edges of the sample remaining below $B_\text{Oe}< 1.5\,$mT for the highest current density investigated, $j = 2.77$\,MAm$^{-2}$. This value is vanishingly small in comparison to the transition fields of the magnetic phases in MnSi, which are between $B=80$\,mT and $B=600$\,mT. It is important to emphasis that the influence of Oersted fields on the absolute field direction increases, when the applied field is small. In this case, the Oersted fields cause a strong relative variation of the applied field strength and direction. Fig.\,\ref{fig:figureS07}\,(b) shows the deviation of the effective field $\boldsymbol{B}_\text{eff} = \boldsymbol{B} + \boldsymbol{B}_\text{Oe}$ from the direction of the applied field~$\boldsymbol{B}$, which points parallel to the vertical $z$-axis, evaluated for the location in the sample, at which the deviation is largest. This effect is strong in the absence of an applied field, whereas applied fields around 50\,mT result in a maximum angle of deviation $\varphi \approx 2^\circ$. This effect may influence measurements in zero field. However, in our study we focus on the magnetic properties of the skyrmion lattice, which is stabilized in magnetic fields between $B=180$\,mT and $B=240$\,mT. In this regime, the effect of Oersted fields are vanishingly small.


\section{Effective model for skyrmion strings}
In this section we develop a phenomenological model for a three-dimensional skyrmion lattice in the presence of a slowly oscillating transverse magnetic field, an applied current, and pinning forces.


\subsection{Energy of skyrmion strings}
Skyrmions in three dimensions are smooth line-like magnetic textures which tend to align parallel to a magnetic field, largely independent of the relative orientation of crystal structure and field direction~\cite{2009:Muhlbauer:Science}. When a weak oscillating transverse field $\boldsymbol{B}_x(t) = b \cos(\omega t) \hat{\boldsymbol{x}}$ is added to a static applied magnetic field $\boldsymbol{B}$ with $b \ll B$, the direction $\hat{\boldsymbol{B}}_{\mathrm{tot}}(t)$ of the total field $\boldsymbol{B}_{\mathrm{tot}}(t) = \boldsymbol{B}_x(t) + \boldsymbol{B}$ changes only slightly, up to a maximum of a few degrees in our experiment, $\alpha \ll 1$. 

The direction of a long skyrmion line of length $L$ cannot follow $\hat{\boldsymbol{B}}_{\mathrm{tot}}(t)\approx \boldsymbol{\hat z} + \frac{\boldsymbol{B}_x(t)}{B}$ instantaneously as this requires skyrmions to move a long distance of order $z\alpha$ with a speed $z\partial_t\alpha$, where $-\frac{L}{2} < z < \frac{L}{2}$ is a coordinate along the skyrmion string. Thus, due to forces arising from damping, gyrocoupling, and pinning (see below), the skyrmion lattice will bend. 

To develop a phenomenological model of this bending effect, we first ignore pinning and describe the position of a skyrmion string by the displacement function
\begin{equation}
	\boldsymbol{u}(z,t) = \left(\!\begin{array}{c} u_x(z,t) \\ u_y(z,t) \\ 0 \end{array}\! \right)
\end{equation}
which only depends on $z$ and the time $t$. We use $\boldsymbol{u}(z,t)$ to describe the (average) distortion of a periodic skyrmion lattice. Importantly, $\boldsymbol{u}(z,t)$ captures  two zero-energy modes of the bulk skyrmion lattice: uniform translations and rotations of the skyrmion lattice but it does not describe inhomogeneous lateral distortions. If we denote the magnetization of the undistorted periodic skyrmion lattice by $\boldsymbol{M}_0(\boldsymbol{r})$, then the distorted skyrmion lattice is approximately described by 
\begin{equation}
	\boldsymbol{ M}(\boldsymbol{r},t) \approx \boldsymbol{ M}_0(\boldsymbol{r}-\boldsymbol{ u}(z,t))+\delta \boldsymbol{ M}(\boldsymbol{ r},t).\label{Mdistorted}
\end{equation}
Importantly, $\boldsymbol{u}$ can be much larger than the typical skyrmion distance and thus $\boldsymbol{M}(\boldsymbol{r},t) - \boldsymbol{M}_0(\boldsymbol{r})$ can be very large while $\delta\boldsymbol{M}$ remains tiny as $b \ll B$. Note that \eqref{Mdistorted} ignores potential changes of the effective distance of skyrmions in the distorted skyrmion lattice. This effect gives, however, only subleading corrections to the energy ($\propto (\partial_z \boldsymbol{u})^4$) and is therefore ignored in the following.

We consider external fields which oscillate with a frequency of up to a few 100\,Hz, three to four orders of magnitude smaller than typical excitation frequencies in the gigahertz regime~\cite{2015:Schwarze:NatMater} and also smaller than the relaxation rates of these modes. This implies that we can formulate an effective theory in terms of $\boldsymbol{u}(z,t)$ only and $\delta\boldsymbol{M}$ in \eqref{Mdistorted} will follow $\boldsymbol{u}(z,t)$ adiabatically.

The change of the free energy per skyrmion line for a weakly distorted skyrmion lattice, $|\partial_z\boldsymbol{u}| \ll 1$, and for a weak oscillating field in the $\hat{\boldsymbol{x}}$ direction, $b \ll B$, is approximately given by
\begin{widetext}
\begin{equation}
	\mathcal{F}=-\frac{\gamma}{2} B_x(t)^2 + \frac{\epsilon_0}{2}\int_{-L/2}^{L/2} dz \;  \left( \left( \partial_z u_x- \frac{B_x(t)}{B'} \right)^2 + \left( \partial_{z} u_y \right)^2\right) +\dots \label{Fskyrmion}
\end{equation}	
\end{widetext}
Here, $B' = B + m_0(N_x - N_z)$ contains a correction from demagnetization factors for non-spherical samples, where $m_0$ is the average magnetization of the skyrmion phase, see Sec.~\ref{App.Epsilon0} below. The equation describes the main physical effect: the skyrmion lowers its energy by aligning parallel to the (internal) magnetic field. Accordingly, the energy minimum is obtained for $\partial_z u_x = \frac{B_x(t)}{B'}$ and $\partial_z u_y=0$. The prefactors $\epsilon_0$ and $\gamma$ in \eqref{Fskyrmion} can be obtained exactly 
from measurements of the magnetization and uniform susceptibilities of skyrmion lattices, as described in Sec.~\ref{App.Epsilon0} below. Above, we omitted anisotropy terms as they are tiny in cubic materials like MnSi as they arise only to higher power in spin--orbit coupling~\cite{2009:Muhlbauer:Science}.

Corrections to \eqref{Fskyrmion} involve either higher powers of $\partial_z \boldsymbol{u}$ or higher derivatives $\partial_z^2\boldsymbol{u}$ which can be ignored for smooth and small deformations of the skyrmion lattice. For example, the bending energy of skyrmion lines is described by $\int dz\, \alpha_0 \left( (\partial_z^2 u_x)^2 + \left(\partial^2_z u_y\right)^2 \right) $ with a prefactor $\alpha_0$ of order $\epsilon_0 R_s^2$, where $R_s$ is the typical skyrmion radius. As we will consider below variations of $\boldsymbol{u}$ occurring on the scale of a fraction of the length of the skyrmion string $L\gg R_s$, such terms may be neglected.

An important consequence of \eqref{Fskyrmion} is that the transverse magnetization $\boldsymbol{m}_x$ (per volume) is obtained from the position of the endpoints of the skyrmion string only
\begin{widetext}
\begin{eqnarray}\label{eq:mperp}
	\boldsymbol{m}_x(t) & = -\frac{1}{V_s} \frac{d\mathcal F}{d \boldsymbol{B}_{x}}=\frac{\gamma}{V_s} \boldsymbol{B}_{x}(t)+\frac{\epsilon_0}{V_s B'} \int_{-L/2}^{L/2} dz\,\partial_z \boldsymbol{ u}(t) \nonumber \\
	&= \chi_x^0 \boldsymbol{B}_{x}(t) +(\chi_\perp^\infty-\chi_\perp^0)B' \,\frac{u_x(L/2,t) - u_x(-L/2,t) }{L}
\end{eqnarray}
\end{widetext}
where we used the results from Sec.~\ref{App.Epsilon0} below. Here, $\chi_\perp^0$ is the transverse susceptibility of the skyrmion phase, which is obtained when the skyrmion lattice remains pinned, $\boldsymbol{u}(\pm L/2)=0$.
The susceptibility
\begin{equation}
	\chi_\perp^\infty=\frac{m_0}{B'}
\end{equation} 
in contrast, is obtained, when the skyrmion lattice perfectly follows the external magnetic field with $u_x= z\,\frac{B_x(t)}{B'}$. This susceptibility will be measured for large amplitudes and small frequencies of the external field. We fit  this value to be $\chi_{\perp}^\infty= 0.25$ to reproduce the measured real part of the susceptibility for large amplitudes, see Fig. 4B and 4D of the main text.  Using $N_x \approx 0.081 $, $N_z\approx 0.61$ for our sample geometry, this corresponds to $m_0=3.9 \cdot10^4$\,A/m or a magnetization of 0.13 $\mu_B$ per formula unit, consistent within 10\% with values reported in the literature \cite{Bauer2012}.


\subsection{Equation of motion and boundary condition}
To describe the dynamics of skyrmions in an oscillating field, we can use the Thiele equation~\cite{Thiele1973} which describes accurately the motion of skyrmions in the presence of weak forces. It is usually derived for the skyrmion coordinate $\boldsymbol{R}(t)$ in a two-dimensional system using the ansatz $\boldsymbol{M}(\boldsymbol{r},t) = \boldsymbol{M}_0(\boldsymbol{r} - \boldsymbol{R}(t))$. Comparing to \eqref{Mdistorted}, for each coordinate $z$ we can identify $\boldsymbol{u}(z,t)$ with a skyrmion coordinate $\boldsymbol{R}(t)$ and obtain the equation of motion for the skyrmion line
\begin{widetext}
\begin{eqnarray}\label{thiele}
	\boldsymbol{\mathcal{G}} \times (\dot{\boldsymbol{u}}(z,t) -\boldsymbol{v_s})+ \mathcal D ( \alpha\dot{\boldsymbol{u}}(z,t) - \beta \boldsymbol{v_s})=-\frac{\delta \mathcal F}{\delta \boldsymbol{ u}(z,t)} + \boldsymbol{ F}^{pin}.
\end{eqnarray}
\end{widetext}
Here, $\boldsymbol{\mathcal{G}}$ is a gyrocoupling vector, given by $\boldsymbol{\mathcal{G}} \approx 4\pi m_s W \hat{\boldsymbol{z}}$ where $m_s$ is the spin density (with units $\hbar$ per volume) and $W = -1$ is the winding number per unit cell~\cite{Thiele1973, 2009:Muhlbauer:Science}. More precisely, $\boldsymbol{{\mathcal G}}$ is oriented parallel to the skyrmion orientation $\boldsymbol{\hat n}$ but we neglect this higher-order effect by setting $\hat{\boldsymbol{n}} = \hat{\boldsymbol{z}} + O(\partial_z\boldsymbol{u})$. Further, $\mathcal{D}$ is the dissipative tensor computed from $\mathcal{D} = \frac{m_s}{2}\int_{UC} d^2 r (\nabla\hat{\boldsymbol{\Omega}})^2$ where $\hat{\boldsymbol{\Omega}}$ is the local orientation of the magnetization and one integrates over the unit cell of the skyrmion lattice.

The presence of an external current gives rise to a spin current with velocity $\boldsymbol{v}_s$ approximately oriented parallel (or antiparallel) to the external current. The two terms proportional to $\boldsymbol{v}_s$ describe the effect of the adiabatic and non-adiabatic spin torques.

When evaluating the elastic force $\frac{\delta\mathcal{F}}{\delta\boldsymbol{u}(z,t)}$ in \eqref{thiele}, it is important to treat the boundary term correctly
\begin{widetext}
\begin{eqnarray}
	-\frac{\delta \mathcal F}{\delta \boldsymbol{ u}}= \epsilon_0  
	\partial_{z}^2\boldsymbol{ u} -\epsilon_0  \left( \delta \left(z - \frac{L}{2}\right)-\delta \left(z + \frac{L}{2}\right)\right) \left(\partial_{z}  \boldsymbol{ u} - \frac{ \boldsymbol{ B}_x(t)}{B'} \right).
\end{eqnarray}
\end{widetext}
Importantly, the magnetic field enters only in the boundary terms at $z=\pm L/2$, consistent with our result that the magnetization depends only on $\boldsymbol{u}(\pm L/2)$, as described in \eqref{eq:mperp}. In the absence of further pinning forces, the slope of the skyrmion line is fixed by $\boldsymbol{B}_x(t)$ in such a way that the skyrmion string is always oriented parallel to $\boldsymbol{B}_{\mathrm{tot}}$ at the surface.


\subsection{Phenomenological slip-stick model for pinning}
Pinning forces from defects in the material substantially complicate the theoretical description. They locally distort the skyrmion lattice such that the function $\boldsymbol{u}(z,t)$ is replaced by a field $\boldsymbol{u}(\boldsymbol{r},t)$ with complex dynamics. Especially the motion of the texture and the nature of the depinning transition is a demanding and largely unsolved problem studied in the field of superconducting vortices, charge density waves, and magnetic skyrmions~\cite{ReichhardtReview2022}. Here, we have a more modest goal. In the spirit of a mean-field theory, we will not keep explicitly track of lateral distortions of the skyrmion lattice but instead focus only on how the disorder affects the {\em averaged} orientation and position of the skyrmion lattice described by $\boldsymbol{u}(z,t)$ only. 

We use a simple phenomenological slip-stick model, similar to Ref.~\cite{2012:Schulz:NaturePhys}, to capture the two main effects from pinning: (i)~a skyrmion does not move when the total force $\boldsymbol{F}$ on the skyrmion arising from external and internal sources is smaller than a critical value, the so-called depinning force $F^p$, and (ii)~a skyrmion moving in the presence of disorder experiences a frictional force oriented antiparallel to the direction of motion of the skyrmion. As a simple approach, we assume that the amplitude of this force is again given by $F^p$ such that
\begin{equation}\label{eq:pinningForces}
	\boldsymbol{ F}^{p} = \left\{\begin{array}{ll} 
		-\boldsymbol{ F} & \text{for } F \le F^p \\[1mm]
		-F^p \frac{\dot{\boldsymbol{u}}}{|\dot{\boldsymbol{u}|}}  & \text{for } F>F^p \end{array} \right. .
\end{equation}
In the pinning regime, $\boldsymbol{F}^p + \boldsymbol{F} = 0 $ and the skyrmion is stuck. Note that the effective pinning force is a highly non-linear implicit function of the forces on the skyrmion and the skyrmion velocity, see Sec.~\ref{numericalImpl} below for the numerical implementation.

This simple model does not capture effects like thermal creep or the correct critical exponents of the depinning transition. Most importantly, it does not capture that skyrmions in some part of the sample may start to move while they are stuck in some other part. Otherwise, it defines a simple-to-use phenomenological model which can be used to understand main experimental features in a semi-quantitative way.

It turns out to be important to distinguish two types of pinning forces, namely those in the bulk and those at the end of the skyrmion line. In the following, these types of pinning forces are referred to as bulk and surface pinning, respectively. If the skyrmion line ends at the surface of the sample, enhanced pinning forces at the surface arise from surface roughness and surface defects. The effect may be even more pronounced when a skyrmion string ends in the bulk of the material. In this case, the magnetic texture forms a Bloch point, as identified with an emergent magnetic monopole in Ref.~\cite{Schuette2014}. Such a singular magnetic configuration may exhibit strong pinning to local defects when compared to the smooth skyrmion textures for which such pinning forces are suppressed by factors of $(a/R_s)^2$, where $a$ is the lattice constant and $R_s$ the skyrmion radius~\cite{mueller2015}. 

Thus we describe the pinning force in \eqref{thiele} by three different terms for the pinning in the bulk and at the top and bottom surfaces
\begin{equation}
	\boldsymbol{ F}^{\mathrm{pin}}= \boldsymbol{F}^{p,b} + \delta (z - L/2) \; \boldsymbol{F}^{p,st} + \delta (z + L/2) \; \boldsymbol{F}^{p,sb},\label{pinningforce}
\end{equation}
where each of the forces follows \eqref{eq:pinningForces} and therefore is a function of $z$ and $t$ depending on applied forces and the local skyrmion velocity. The forces are parametrized by three constants, namely $F^{p,b}$, $F^{p,st}$, and $F^{p,sb}$. Note that the bulk parameter $F^{p,b}$ is given in a unit of force per length, while the surface terms are given in a unit of force. 


\subsection{Rescaled equation of motion}
It is instructive to rewrite the Thiele equation through the rescaled dimensionless variables $\tilde{u} = \tilde{u}(\tilde z,\tilde t) = \epsilon_0 u(z,t) / (F^{p,b} L^2)$, $\tilde{z} = \frac{z}{L}$, and $\tilde{t} = \frac{\epsilon_0 t}{L^2 \mathcal{G}}$ with
\begin{widetext}
\begin{eqnarray}\label{thieleRescaled}
	& \hat{\boldsymbol{z}} \times \dot{\tilde{\boldsymbol{ u}}}+  \frac{\alpha \mathcal D}{\mathcal{G}} \dot{\tilde{\boldsymbol{ u}}} =
	\boldsymbol{v}_s^{\mathrm{eff}}+\partial_{\tilde{z}}^2 \boldsymbol{\tilde{u}}- \left(\delta \!\left( \!\tilde{z} - \frac{1}{2}\!\right)-\delta\! \left( \! \tilde{z} + \frac{1}{2}\!\right)\right) \; \left(\partial_{\tilde{z}} \boldsymbol{\tilde{u}} - \delta \tilde{B} \hat{\boldsymbol{x}}\right) +\tilde{\boldsymbol{F}}^{\mathrm{pin}},\\
	&\tilde{\boldsymbol{F}}^{\mathrm{pin}} =\frac{\boldsymbol{F}^{p,b}}{F^{p,b}}+
	\delta (\tilde{z} + 1/2) \; \frac{\boldsymbol{F}^{p,st}}{L F^{p,b}} + \delta (\tilde{z} - 1/2) \;\frac{\boldsymbol{F}^{p,sb}}{L F^{p,b}} ,\nonumber
\end{eqnarray}
\end{widetext}
where we introduced a dimensionless effective spin current velocity and rescaled angle of magnetic field
\begin{equation}
	\boldsymbol{v}_s^{\rm eff}=\frac{1}{F^{p,b}} \left( \mathcal{G}\times  \boldsymbol{v}_s+
	\beta\mathcal D  \boldsymbol{v}_s\right), \; \delta \tilde{B} = \frac{\epsilon_0}{L F^{p,b}}\frac{b}{B'}.\label{eq:rescaled}
\end{equation}

The external current controls the amplitude of $\boldsymbol{v}_s^{\mathrm{eff}}$ and the external transverse field determines the strength of the magnetic field $\delta\tilde{B}(\tilde{t}) = \delta\tilde{B}\cos(\tilde{\omega}\tilde{t})$, where the dimensionless frequency $\tilde{\omega}$ is given by $\omega/\omega_0$, where $\omega_0 =  \frac{\epsilon_0}{L^2 \mathcal{G}}$ is a characteristic frequency of tilting skyrmion lattices which can become very small for large $L$. The only remaining unknown parameter is the ratio of surface to bulk pinning, $F^{p,st}/(L F^{pb})$ and $F^{p,sb}/(L F^{pb})$, which turns out to be of the order $1$ from fits to the experiments, see below.

The complex transverse susceptibility, $\chi_\perp(\omega)$, in a response to an oscillating transverse field $B_x(t) = b\cos(\omega t)$, which is equivalent to $\delta \tilde{B}(t) = \delta\tilde{B}\cos(\tilde{\omega}\tilde{t})$ in rescaled coordinates, is computed by integrating the transverse magnetization over one oscillation period $T = 2\pi/\omega$ using 
\begin{widetext}
\begin{align}
	\chi_{\bot}(\omega) &=  \frac{2}{T b}\int_{t_0}^{t_0+T} m^x_\perp(t) e^{i \omega t} \,dt \nonumber \\
	&=\chi^0_\perp+(\chi^\infty_\perp-\chi^0_\perp) \frac{2}{\tilde T }
	\int_{\tilde t_0}^{\tilde t_0+ \tilde T} 
	\frac{\tilde{u}_x(1/2,\tilde t) - \tilde{ u}_x(-1/2,\tilde t)}{\delta \tilde{B}}  e^{i \tilde \omega \tilde t} d\tilde t.
\end{align}
\end{widetext}
Here, $t_0$ is an initial time chosen sufficiently large (typically $t_0 = 20 T$ in our implementation) such that all initial state effects have vanished. In the linear response regime, when the skyrmion lattice is always pinned, one recovers the transverse susceptibility $\chi_\perp^0$, while $\chi_\perp^\infty$ is obtained when the skyrmion lattice follows the external magnetization, $\tilde{u}_x = z\,\delta\tilde{B}(\tilde{t})$.


\subsection{Numerical implementation}
\label{numericalImpl}
Due to the presence of pinning, the Thiele equation of the dynamics of the skyrmion line, given in \eqref{thieleRescaled}, is a highly non-linear two-dimensional differential equation that was solved numerically. 

To determine the dynamical properties we discretized the skyrmion line both in space and time. We replaced the line in $z$ direction ($z \in \{-\frac{1}{2},\frac{1}{2}\}$) by $N$ points with a distance $\frac{1}{N-1}$ between each point, with one point at the beginning and one at the end of the line. Time was discretized, such that during each period $T_{\mathrm{period}} = \frac{2\pi}{\tilde{\omega}}$ of the oscillation of the applied magnetic field $N_{\mathrm{steps}}$ updates were performed, with a time step of $d\tilde{t} = \frac{T_{\mathrm{period}}}{N_{\mathrm{steps}}}$. This way, the problem of determining the dynamics effectively reduced to finding coordinates of $N$ points at every time step. For the integration of \eqref{thieleRescaled} Heun's method of finding numerical solutions to ordinary differential equations was used with the starting condition of a straight line (zero displacement of every point).

Given the positions of all points $u_i^j$  ($i=1\dots N$) at time step $j$, the elastic force acting on each point can be calculated by means of the first and the second numerical spatial derivative for the first and the last point, and all intermediate points, respectively. In our code, this approach was implemented using the second-order centered difference approximation, which represents the error when calculating derivatives of the order $\sim \left(\frac{1}{N}\right)^2$.

As the first step, we checked for every point if the total force acting on it was smaller than the pinning force, given by $(N-1) \tilde{F}^{p,st}$ for the first point, $\tilde{F}^{p,b}$ for points with $j=2 \dots N-1$, and $(N-1) \tilde{F}^{p,sb}$ for the last pointy. Provided the total force was found to be smaller then the pinning force, the point was kept unchanged without update. If the total force on the point was found to be larger than the pinning force, we solved the resulting \emph{nonlinear} equation for the velocity $v_i=\dot{\tilde{\boldsymbol{u}}}_i$ related to the $1/|\partial_t \tilde{\boldsymbol{ u}}_i|$ term in \eqref{pinningforce}. This equation can be solved analytically (not shown) and the resulting velocity is used within Heun's method with a truncation error of the order $\sim \left(\frac{1}{N_{\mathrm{steps}}}\right)^2$.

In our simulations, we typically use $N_{\mathrm{steps}} = 3\cdot10^5$ and $N=60$ while waiting for around 20 periods to obtain a periodic solution in the long-time limit.


\subsection{Phase diagrams of the effective model}

\begin{figure}[htb!]
	\centering
	\includegraphics[width=1\linewidth]{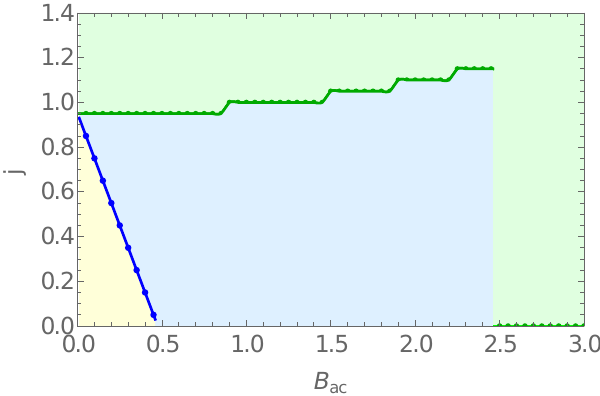}
	\caption{Calculated phase diagram in the absence of surface pinning using parameters $\omega/\omega_0 = 2.32$, $\alpha \mathcal D/\mathcal{G} = 0.1$, $F^{p,sb}=F^{p,st}=0$, and $\beta \mathcal D/G=0.07$. The following phases are distinguished: partially pinned (yellow shading), walking (blue shading), and running (green shading).}
	\label{fig:figureS08}
\end{figure}

\paragraph*{No surface pinning}
Without surface pinning, an arbitrarily weak oscillation of the magnetic field leads to a motion of the ends of the skyrmion string, while the center remains pinned. This partially pinned phase for small currents and fields is visualized in Fig.\,1B of the main text. The corresponding phase diagram is shown in Fig.~\ref{fig:figureS08} in which this partially pinned phase is accompanied by the walking phase and the running phase. Compared to the phase diagram shown in Fig.\,1 of the main text, a key difference concerns the transition from the walking to the running phase which is observed even for vanishing current density. Notably, under large oscillation amplitudes, the skyrmion line oscillates in both lateral directions without ever being pinned by disorder.  

\begin{figure}[htb!]
	\centering
	\includegraphics[width=1\linewidth]{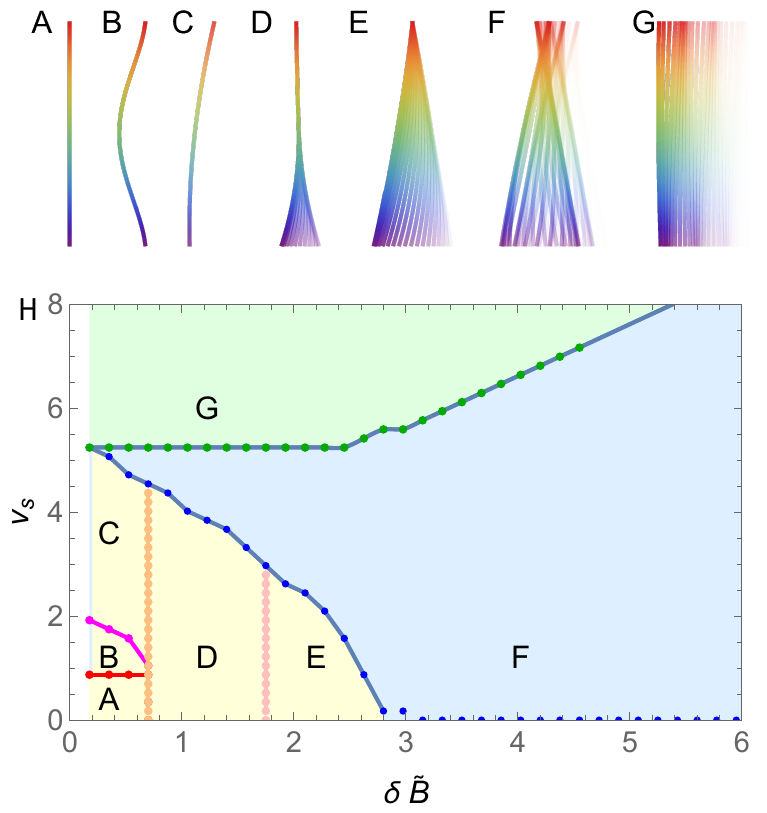}
	\caption{Calculated phase diagram for asymmetric surface pinning. \mbox{({A--G})}~Visualization of the bending and motion of skyrmion lines for stroboscopic times as illustrated by decreasing color saturation. Seven qualitatively different phases may be distinguished. Characteristic configurations $u_x(t)$ are shown for stroboscopic times $t = n\delta t$. (H)~Calculated phase diagram using parameters $\omega/\omega_0 = 2.32$, $\alpha \mathcal D/\mathcal{G} = 0.1$, $F^{p,sb}/(L F^{p,b})=0.7,F^{p,st}/(L F^{p,b})=3.5$, and $\beta \mathcal D/G=0.07$. The position of the capital letters denotes the values of $\delta\tilde{B}$ and $v_s$ visualized in panels ({A--G}). The pinned phases (yellow shading) may be divided into fully pinned phases (A, B, and C) and partially pinned phases (D and E). See text for details.}
	\label{fig:figureS09}
\end{figure}

\paragraph*{Isotropic surface pinning}
Fig.\,1 of the main text shows the phase diagram for an isotropic surface pinning comprising the fully pinned phase, the partially pinned phase, the walking phase, and the running phase. In fact, in the fully pinned phase, two scenarios may be distinguished. For $j<0.57$, the trivial solution is obtained as shown in Fig.\,1 of the main text. For $j>0.57$, a time-independent curved solution is obtained, as shown in Fig.\,\ref{fig:figureS09}\,B. The ends of the line are pinned, while the center of the skyrmion line is already depinned due to the current. In both scenarios, the skyrmion line remains static and no response is observed in the magnetic susceptibility.


\begin{figure}[htb!]
	\centering
	\includegraphics[width=1\linewidth]{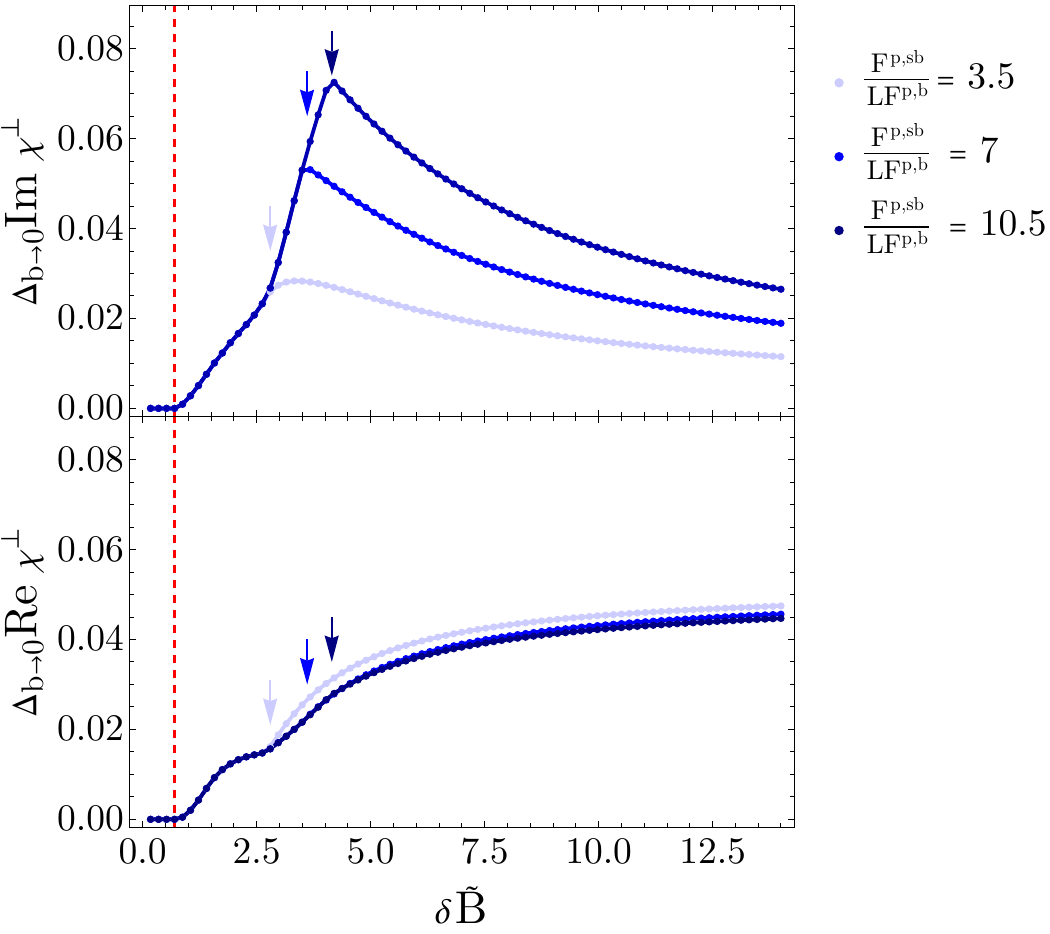}
	\caption{Change of the real and imaginary parts of the calculated susceptibility as function of the amplitude of the oscillating field $\delta\tilde{B}$ for a different strength of anisotropic surface pinning. Calculations were performed using parameters $\omega/\omega_0 = 2.32$, $\alpha \mathcal D/\mathcal{G} = 0.1$, $F^{p,st}/(L F^{p,b})=0.7$, $\beta \mathcal D/G=0.07$). 
		The red dashed line denotes the critical value of the field where the transition from the fully to the partially pinned phase takes place (Fig.\,\ref{fig:figureS09} A$\rightarrow$ D), while the arrows mark the critical field value of the transition from the partially pinned to the walking phase (Fig.\,\ref{fig:figureS09} E$\rightarrow $F). The colors of the arrows encode the corresponding values of anisotropic surface pinning.}
	\label{fig:figureS10}
\end{figure}

\begin{figure}[htb!]
	\centering
	\includegraphics[width=1\linewidth]{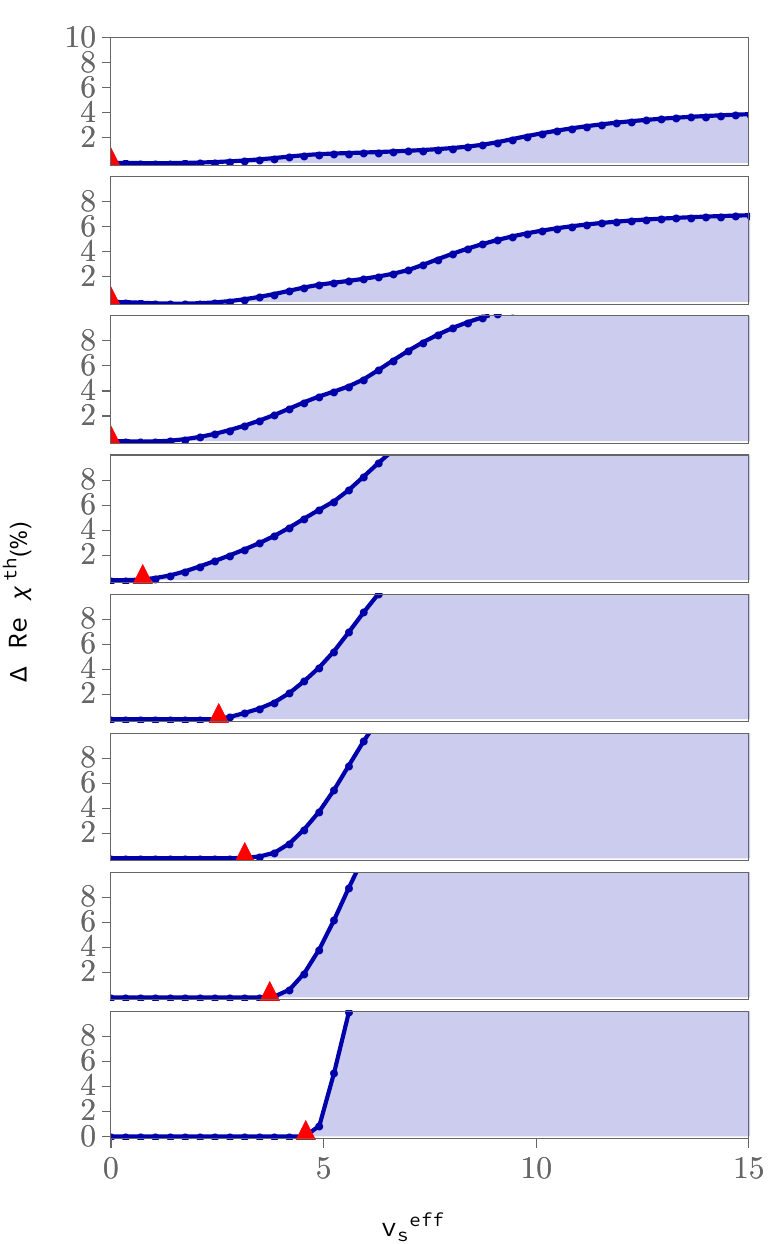}
	\caption{Relative change of the real part of the calculated transverse susceptibility for asymmetric surface pinning as a function current density $j$ for different $\delta\tilde{B}$. Parameters: $\omega/\omega_0 = 2.32$, $\alpha \mathcal D/\mathcal{G} = 0.1$, $F^{p,sb}/(L F^{p,b})=0.7$, $F^{p,st}/(L F^{p,b})=3.5$, and $\beta \mathcal D/G=0.07$.	}
	\label{fig:figureS11}
\end{figure}

\paragraph*{Anisotropic surface pinning}
In general, the pinning at both ends of a skyrmion string may be different. Thin-film samples grown on a substrate represent an obvious example for which the pinning at the top and at the bottom surface of a sample may differ strongly. Moreover, when skyrmion strings are shorter than the size of the sample, one end of the skyrmion string may be located at the surface of the sample while the other end is located in the bulk, potentially giving rise to distinctly different pinning forces at the ends of the string.

In Fig.\,\ref{fig:figureS09}, we consider such a situation where the surface pinning is asymmetric. For the parameters chosen, we observe five fully pinned or partially pinned phases which differ in terms of which parts of the skyrmion string are deformed or moving, respectively. In the three fully pinned phases, the skyrmions are stationary in the long-time limit and distinguished by their static response only. In phase A, no changes are observed in the shape of the skyrmion line. In phase B, the finite current bends the skyrmion line, but the ends of the line remain pinned. In phase C, one end of the line remains pinned while the other is depinned, leading to a change in magnetization. The two partially pinned phases are distinguished by which part of the skyrmion string is moving. In phase D, a finite fraction of the line remains fully pinned. In phase E, the entire line oscillates without net motion in space as one of the ends remains pinned. In the walking phase (F) and the running phase (G), the skyrmion lines exhibit a net motion, similar to the situation under symmetric pinning.

In Fig.\,\ref{fig:figureS10}, we illustrate how such anisotropic surface pinning affect the dependence of the real and imaginary parts of the susceptibility on the oscillation amplitude $\delta\tilde{B}$. In the real part, an intermediate plateau emerges at small oscillation amplitudes. The imaginary part drastically increases as the pinning becomes more anisotropic, exhibiting a maximum at an oscillation amplitude slightly above the transition from the partially pinned to the walking phase. It is interesting to speculate that the pronounced maximum observed in $\mathrm{Im}\chi_\perp$ in our experiments may reflect anisotropic pinning in the presence of a broad distribution of pinning forces. 

In Fig.\,\ref{fig:figureS11}, we show how the asymmetric pinning affects the current dependence of $\mathrm{Re}\chi_\perp$. When comparing these calculation with the results obtained under symmetric surface pinning, shown in Fig.\,4 of the main text, both qualitatively and quantitatively similar behavior is observed.


\subsection{Computation of $\epsilon_0$ and $\gamma$}
\label{App.Epsilon0}
The prefactors $\epsilon_0$ and $\gamma$ in \eqref{Fskyrmion} can be computed exactly from the free energy density $f$ of skyrmion lattices oriented parallel to a vector $\hat{\boldsymbol{n}}$ in a fixed magnetic field $\boldsymbol{B}$. The uniform magnetization per volume is written in a Taylor expansion around $\boldsymbol{m}_0 = m_0\hat{\boldsymbol{n}}$. Here, $m_0$ is the average magnetization density of skyrmions in the internal background field $B_0$, corresponding to the external field $B = B_0 + N_z m_0$ when the demagnetization factor $N_z$ is taken into account, oriented parallel to $\hat{\boldsymbol{n}}$. The Taylor expansion of the free energy is given by 
\begin{widetext}
\begin{equation}
	f=\frac{d f_0}{d \boldsymbol{ m}} \cdot (\boldsymbol{ m}-m_0 \boldsymbol{\hat n})+\frac{1}{2 \chi^{00}_\|} (\boldsymbol{m}-m_0 \boldsymbol{\hat n})^2+\frac{1}{2 \chi^{00}_\perp} (\boldsymbol{ m}_\perp)^2+ \sum_i \frac{N_i m_i^2}{2} - \boldsymbol{ m} \boldsymbol{ B},
\end{equation}
\end{widetext}
where $f_0$ is the free energy density in the absence of both an external field and demagnetization factors, $N_i$ are the demagnetization factors encoding the sample shape, and $\chi^{00}_{\perp/\|}$ are the bare (static) susceptibilities of the skyrmion lattice for a transverse/parallel magnetic field (in the absence of demagnetization fields). Using $\frac{d f_0}{d \boldsymbol{ m}}=-B_0 \boldsymbol{\hat n}$ and identifying $\mathcal{F} = \min_{\boldsymbol{m}} F$, we obtain for $N_x = N_y = N_\perp$ the following exact result
\begin{eqnarray}
	\epsilon_0&=A_s (\chi_\perp^\infty-\chi_\perp^0) {B'}^2 \\
	\gamma&=\chi_\perp^\infty V_s,
\end{eqnarray}
where $A_s\approx 230\,\text{nm}^2$ is the area per skyrmion, $V_s = A_s L$ the corresponding volume, and $\chi_\perp^0 = \frac{\chi_\perp^{00}}{1+N_x \chi_\perp^{00}}$ is the susceptibility in the pinned phase. The susceptibility 
\begin{equation}
	\chi_\perp^\infty=\frac{m_0}{B'}=\frac{m_0}{B+m_0 (N_x-N_z)}\\
\end{equation}
is observed for large amplitudes and small frequencies, when the pinning can be neglected such that $\hat{\boldsymbol{n}}$ follows the applied magnetic field and can be computed from the magnetization. From the experimental data, we obtain the estimate $\chi_\perp^0 \approx 0.2/ \mu_0$ where $\mu_0$ is the magnetic permeability, while $\chi_\perp^\infty = \frac{m_0}{B'} \approx 
0.25/\mu_0$. 

From these values, we obtain
\begin{equation}
	\epsilon_0 \approx 4\cdot 10^{-13} \frac{\text{J}}{\text{m}}\approx 30 \, \text{K} \frac{k_B}{\text{nm}}\approx 600 \, \text{GHz}  \frac{2 \pi \hbar}{\text{nm}}\label{epsilon_0}.
\end{equation}


\subsection{Length and energy scales induced by pinning and the breakup of skyrmion lines}
\label{App.Scales}
In this section, we discuss important length and energy scales governing the response of the skyrmion to an oscillating magnetic field. In the presence of a bulk pinning potential, an oscillating skyrmion string starts to bend. The bending radius, $L_p$, is obtained by balancing the elastic force, $\epsilon_0 \partial_z^2 u$ with the pinning force $F^p$, leading to 
\begin{equation}
	L_p \approx \frac{\epsilon_0}{F^p}.
\end{equation}
As pinning in our system is very weak, $L_p$ is very large. Ignoring surface pinning, we can estimate $L_p$ when using that the skyrmion bends by an angle $b/B$ at the critical depinning field,
\begin{equation}
	\frac{L}{L_p} \sim \frac{b}{B} \sim 0.02 \ll 1 \label{Lp}.
\end{equation}
Thus, in the regime considered in this study, $L_p$ is much larger than the typical size of the skyrmion strings.

An alternative estimate of the pinning forces and therefore of $L_p$ may be inferred from the critical current for depinning, which is of the order of $10^6\,\mathrm{Am}^{-2}$. As described in Ref.~\cite{2012:Schulz:NaturePhys}, this value corresponds to a drift velocity of the order of $v_s\approx 10^{-4}$\,m/s. Therefore, we estimate the pinning force $F^p \approx \mathcal{G} v_s \approx 4\pi\frac{m_0}{\gamma_e} v_s $ with the gyromagnetic ratio $\gamma_e$ and 
$m_0\approx 4.4\cdot10^4$\,A/m. Combining this result with our estimate for $\epsilon_0$, \eqref{epsilon_0}, we obtain the order-of-magnitude estimate 
\begin{equation}
	L_p \sim 1-10\,\text{mm}.
\end{equation}
where the lower boundary stems from the arguments given above while the upper boundary takes into account that inelastic spin-flip scattering at finite temperatures in MnSi reduces the topological Hall effect and therefore also $\mathcal{G}$ by an order of magnitude compared to the value used above~\cite{Ritz2013}. This uncertainty makes definite conclusions on the value of $L$ challenging. Plausible values of $L$ may be between $15\,\mu$m and the sample height of $400\,\mu$m.

A central question concerns whether the bending of skyrmions induced by the oscillating field can lead to a `breaking-up' of the skyrmion string into smaller pieces. To estimate this effect, we use again $\epsilon_0 \partial_z^2 u \sim F_p = \epsilon_0/L_P$ and therefore $u\sim z^2/L_p$, leading to a total bending energy of $\epsilon_0 \frac{L^3}{L_p^2} \approx \epsilon_0 L\, 4\cdot 10^{-4}$. The energy scale needed to break a skyrmion line is set by the energy to create a pair of Bloch points (a monopole--antimonopole pair)~\cite{Wild2017}. A reliable estimate of this energy scale is difficult for a system like MnSi, which is governed by thermal fluctuations~\cite{2009:Muhlbauer:Science}. Experiments on the lifetime of skyrmion strings show that it is strongly renormalized by entropic contributions~\cite{Wild2017} and considerably smaller than estimates based on the energetics alone. Ignoring this effect and estimating the skyrmion-breaking energy to be $T_c R_s/a$~\cite{Wild2017}, where $T_c$ is the transition temperature and $R_s/a\sim 10^2$ is the ratio of skyrmion radius and lattice constant, we find that skyrmions may break for $\epsilon_0 \frac{L^3}{L_p^2} \gtrsim T_c \frac{R_s}{a}$. This consideration translates into
\begin{equation}
	L \gtrsim \left( \frac{T_c R_s L_p^2}{\epsilon_0 a} \right)^{1/3}\sim 0.05 - 0.2\, \text{mm}
\end{equation}
or shorter if our estimate for the energy scale is too big.

Our order-of-magnitude estimates are consistent with two different scenarios: (i) the length of skyrmion strings is set by the system size, or (ii) the interplay of pinning and the bending of skyrmion lines by an oscillating field breaks the skyrmion into smaller pieces. The latter effect is not covered by our theoretical treatment but may explain why in our experiments the change of susceptibility induced by an oscillating field is much larger than that induced by a current. Smaller skyrmion pieces can follow the oscillating magnetic field more easily. This effect can, for instance, be seen in \eqref{eq:rescaled}, where the dimensionless oscillating field grows when $L$ is reduced.

